\documentclass[sigconf,screen,authorversion,nonacm]{acmart}
\usepackage{fancyhdr}
\AtBeginDocument{%
  
      \addtolength{\footskip}{2.0\baselineskip}%
    \fancyfoot[L]{\textit{\textbf{Preprint --- Author version September 2023.}}}%
    }

\setcopyright{acmcopyright}
\copyrightyear{2018}
\acmYear{2018}
\acmDOI{XXXXXXX.XXXXXXX}

\acmConference[Authors version]{}{Preprint}{September 2023}
\acmPrice{15.00}
\acmISBN{978-1-4503-XXXX-X/18/06}

\newcommand{\m}[1]{{\fontfamily{mathptmx}\selectfont\textit{#1}}}

\newcommand{\defgroup}{{\m{Group}}}
\newcommand{\defcongruency}{{\m{Congruency}}}

\newcommand{\defsitusitu}{{\m{Situ$\rightarrow$Situ}}}
\newcommand{\defsituNONsitu}{{\m{Situ$\rightarrow$NonSitu}}}
\newcommand{\defNONsitusitu}{{\m{NonSitu$\rightarrow$Situ}}}
\newcommand{\defNONsituNONsitu}{{\m{NonSitu$\rightarrow$NonSitu}}}
\newcommand{\defSitu}{{\m{Situ}}}
\newcommand{\defNonSitu}{{\m{NonSitu}}}

\newcommand{\defLearning}{{\m{Learning}}}
\newcommand{\defRecall}{{\m{Recall}}}

\newcommand{\defModality}{{\m{Modality}}}
\newcommand{\defDesktopOne}{{\m{Desktop 1}}}
\newcommand{\defDesktopTwo}{{\m{Desktop 2}}}
\newcommand{\defVR}{{\m{VR}}}

\definecolor{color1}{RGB}{230,97,1}
\definecolor{color2}{RGB}{253,184,99}
\definecolor{color3}{RGB}{94,60,153}
\definecolor{color4}{RGB}{178,171,210}

\def\defattributeone#1{\textcolor{color1}{\textbf{#1}}}
\def\defpatternone#1{\textcolor{color2}{\textbf{#1}}}
\def\defattributetwo#1{\textcolor{color3}{\textbf{#1}}}
\def\defpatterntwo#1{\textcolor{color4}{\textbf{#1}}}
\usepackage{tabularray}
\usepackage{tabu}
\usepackage{enumitem}
\begin{document}

\title{
Context-Dependent Memory in Situated Visualization



}

\author{Kadek Ananta Satriadi}
\affiliation{%
  \institution{Monash University}
  \city{Melbourne}
  \country{Australia}
}

\author{Benjamin Tag}
\affiliation{%
  \institution{Monash University}
  \city{Melbourne}
  \country{Australia}
}

\author{Tim Dwyer}
\affiliation{%
  \institution{Monash University}
  \city{Melbourne}
  \country{Australia}
}

\renewcommand{\shortauthors}{Satriadi et al.}

\begin{abstract}
 Situated visualization presents data alongside their source context (physical referent). While environmental factors influence memory recall (known as Context-Dependent Memory or CDM), how physical context affects cognition in real-world tasks such as working with visualizations in situated contexts is unclear. This study explores the design space of information memorability in situated visualization through the lens of CDM. We investigate the presence of physical referents for creating contextual cues in desktop and Virtual Reality (VR) environments. Across three studies ($n=144$), we observe a trend suggesting a CDM effect due to contextual referent is more apparent in VR. 
Overall, we did not find statistically significant evidence of a CDM effect due to the presence of a referent. However, we did find a significant CDM effect for lighting condition.
This suggests that representing the entire environment, rather than the physical objects alone, may be necessary to provide sufficiently strong contextual memory cues. 
\end{abstract}



\received{20 February 2007}
\received[revised]{12 March 2009}
\received[accepted]{5 June 2009}

\begin{teaserfigure}
\includegraphics[width=\textwidth]{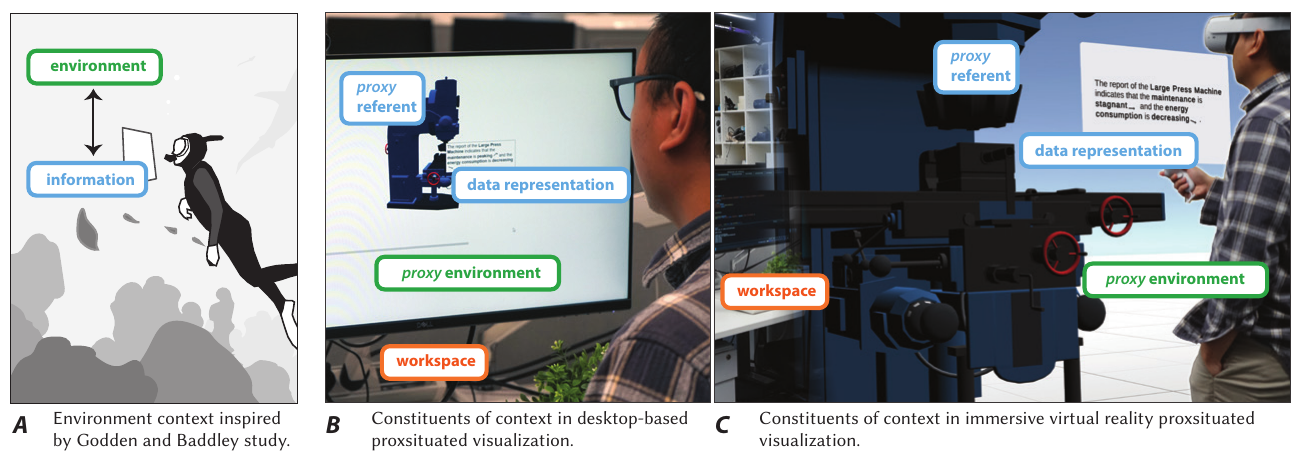}
\caption{A) Context-dependent memory studies have investigated the effect of learning and recall environments on memory recall performance. A similar concept can be applied to situated visualization but what constitutes context can go beyond the environment that surrounds the user. 
In \textit{proxsituated} scenarios such as desktop-based (B) and immersive virtual reality (C) visualizations with digital models, there are complex interactions between the workspace, proxy environment, proxy referent, and data representation. Where can we draw a fine boundary of contextual memory cues within these factors?}
\label{fig:teaser}
\Description{The figure has three parts. The first part shows an underwater illustration with a diver reading learning material. The second part shows a desktop-based situated visualization displaying an information panel and a 3D model of an industrial machine. The third part depicts the immersive virtual reality version of the second image.}
\end{teaserfigure}

\maketitle

\section{Introduction}

Situated visualization assumes that the presence of a physical referent (the object that is the source or subject of data being visualized) and its environment is important to real-world ``in-situ" data analysis workflows.  There are pragmatic benefits to making information available when and where it is needed through wearable devices, for example making it possible to make data-informed decisions while in the field or in the physical workplace rather than having to return to an office.  
But proponents have also argued that there are cognitive benefits to bringing data into the world, based on people's natural spatial abilities and theories of embodied action, e.g. \cite{thomas2018situated}.  
However, efforts to measure or quantify the supposed benefits of such embodiment are in their infancy.
In this paper, we investigate one fairly well-known cognitive dimension involving the effect of environment on memory and seek to understand whether it plays a role in situated visualization. That is, we aim to extend our understanding of the relationship between data representation and physical referent by investigating the impact of the presence of the physical referent on the chart's memorability. In particular, we investigate whether context-dependent memory has a role in situated visualization. 

Context-dependent memory (CDM) theory suggests that the learning and recall contexts influence the accuracy of information recall. As stated by ~\citet[p.112]{smith2007context}, ``\textit{Memory is said to be context-dependent because experiences always occur within context, and memories of events depend upon the contexts in which those experiences occur}''.
In 1975, \citet{godden1975context} conducted a seminal study comparing learning and recall across two distinct contexts: either underwater wearing scuba apparatus, or on dry-land.  Specifically, subjects learned and recalled words in four conditions: land-land, land-water, water-land, and water-water (illustrated in Figure \ref{fig:teaser}-A). They discovered that recall performance was superior when learning and recalling occurred in the same contexts (land-land, water-water) compared to different contexts (land-water, water-land). While the CDM study offers a fresh perspective on situated visualization, there has been no prior effort to bridge the gap between these two fields.

Studies in information visualization have shown that incorporating contextual information, including so-called 'chart junk' or decorative images with charts, can enhance their long-term memorability~\cite{bateman2010useful}. Additionally, previous research has indicated that the inclusion of familiar elements like objects, humans, and scenes can improve memorability~\cite{borkin2013makes}. However, it remains unexplored whether these findings can be attributed to CDM. To the best of our knowledge, there has been no data visualization study employing CDM experimental designs. Consequently, the potential role of physical referents as contextual cues from a CDM perspective has not been empirically investigated. We provide an overview of related concepts in Section \ref{sec:background} in this paper. 

This paper contributes in two main ways. Firstly, it outlines a design space for exploring CDM in situated visualization (Section \ref{designspace}). This encompasses a variety of factors, such as the type of physical referent (environment, single referent, homogeneous referents, and heterogeneous referents), and different forms of situatedness (immediate situated, prox-situated in non-VR settings, and VR prox-situated), among others. 
Secondly, we present the findings from three studies inspired by this design space (Section \ref{studies}). These studies investigate the impact of the presence of physical referent representations in prox-situated scenarios, both on desktop (Figure \ref{fig:teaser}-B) and in VR (Figure \ref{fig:teaser}-C). The results (Section \ref{results}) indicate that while there was some effect of physical referent presence on memory accuracy, these effects were not statistically significant on their own. Instead, the studies suggest that factors like the environment and workspace in desktop settings, as well as the immersive environment surrounding the referent in VR settings, may play more influential roles in inducing CDM effects. We further contextualize the results and reflect on situated visualization (Section \ref{discussion}), then acknowledge limitations and suggest directions for future work (Section \ref{limitations}).

\section{Background}
\label{sec:background}
In this section, we provide an overview of context-dependent memory theory and the state of the research landscape on situated visualization, as well as memory-related studies in information visualization.

\subsection{Context-Dependent Memory}

Researchers in psychology, neurobiology, philosophy, education, and computer science use different descriptions and methodologies to define \emph{memory}~\cite{dudai2007memory}. A common factor across the wide variety of concepts is the notion of \emph{context}. 
``\textit{Context refers to the situation or circumstances in which an event takes place}'', ~\citet[p.99]{roediger2007science}. 
 In Cognitive Psychology, it is argued that contextual features can link memory to associated events~\cite{godden1975context}. In their study, \citet{godden1975context} found that the performance of recall of people who switch their learning and recall environments (e.g., learning underwater and recalling on land) is outperformed by that of those who did not switch their environments, suggesting context-item integration~\cite{smith2007context} between learned information and the features of the environment in one's mind.  
Since then, research around Context-Dependent Memory (CDM) has flourished, attempting to determine its predictability and quantify the phenomenon. 

As \citet{smith1994theoretical, smith2001environmental} describe in their analyses, context is formed by multiple features that act as contextual cues. These cues can create contextual binding or association with target information and stimulate the memory of that information. Contextual cues can come internally from the person's own state (e.g., mood) or externally from the environment. The binding between cues and targets can be through incidental learning or intentional learning. When it comes to remembering the information, the contextual cues memorized during learning can not only come from the immediate environment but also from a mental representation of the environment in the person's mind: remembering context can help remember the target information.

The literature also offers an introduction to the typical research paradigms employed to deduce the CDM effect. These paradigms encompass context reinstatement, interference reduction, and multiple-context. In our research, we place particular emphasis on the reinstatement paradigm, which posits that recalling the learning context can enhance the recall of learned material. For more details on these concepts, including overshadowing and outshining, please refer to Appendix \ref{appendixcdm}.

\subsection{Recent CDM Studies}

The CDM concept has not yet been a core focus of Human-Computer Interaction research. Therefore, we summarize recent studies with HCI relations in the following section (see Table~\ref{table:lit}, in the Appendix \ref{appendixrelatedwork} for the list of papers).  

Various types of contextual environments have been explored in these studies. These environments encompass desktop scenes, including pictures of sceneries, color, and flickering screens~\cite{walti2019reinstating}. Researchers have also delved into immersive environments. As demonstrated by deBack, early work involved the use of panoramic images on a CAVE display~\cite{de2018applicability}. More recent studies have leveraged virtual reality headsets, featuring environments like Mars~\cite{shin2021context}, underwater settings~\cite{parker2020exploring, shin2021context}, replicas of real-world spaces~\cite{chocholavckova2023context, mizuho2023effects, lamers2021changing}, and fantasy worlds~\cite{essoe2022enhancing}. Most studies have compared two different contextual environments within the same modality, such as desktop-only~\cite{walti2019reinstating} or VR-only~\cite{walti2019reinstating, shin2021context, essoe2022enhancing, chocholavckova2023context}, while a few have compared multimodal contexts, such as virtual versus real physical environments~\cite{mizuho2023effects, lamers2021changing, parker2020exploring}. From this array of studies, we summarize the relevant key trends.

Multiple studies have found that context benefits memory in the long term~\cite{essoe2022enhancing, shin2021context, lamers2021changing}. \citet{shin2021context}, evaluated two different immersive virtual environments, planet Mars and underwater, in VR. Their study found a CDM effect and suggested that the effect is delayed. This means the ability to recall information is not immediately available after information processing but is delayed by 24 hours. This finding is indirectly corroborated by other studies that did not find a significant CDM effect when investigating short retention~\cite{walti2019reinstating, parker2020exploring,chocholavckova2023context, mizuho2023virtual}. The only exception is \citet{de2018applicability}, which used panoramic mountain and underwater scenes on a CAVE display. They found differences between the Mountain-Mountain and Mountain-Underwater conditions when testing short-term retention. However, there was no significant difference between the Underwater-Underwater and Underwater-Mountain conditions. They explained that the underwater scene, presented exclusively during recall, had a more significant impact due to its novelty and acted as a distraction.

High presence (the sense of ``being there'' in a virtual environment) appears to be advantageous for context-dependent memory (CDM), while the increased realism of environments may not be as beneficial, as noted in prior research~\cite{essoe2022enhancing, shin2021context, lamers2021changing}.
For instance, \citet{mizuho2023effects} examined real, high-fidelity, and low-fidelity virtual environments. They found that while enhanced visual fidelity contributed to a stronger sense of presence, it did not significantly impact participants' ability to recall a list of words.
In a recent study, \citet{essoe2022enhancing} affirmed that a high level of presence, a characteristic of VR, benefits CDM. Additionally, they highlighted that a unique context demonstrates the enduring advantages of CDM, even over longer periods, such as one week.

It's worth noting that there are other memory-related studies not directly related to context-dependent memory but still relevant to the effect of realism. \citet{harman2019role}, for example, found no discernible effect of visual fidelity in virtual environments (high vs. low) on recognition tests for objects. Similarly, \citet{lee2013effects} investigated the impact of transitioning from the physical world to virtual ones with varying levels of visual fidelity on the ability to locate objects within the scene but found no significant effects. Conversely, a study exploring the transferability of navigation skills from virtual to physical environments demonstrated the benefits of increased visual fidelity~\cite{wallet2011virtual}.

\subsection{Situated Visualization}
Situated visualization has a research history of nearly two decades, with the term being introduced in 2006 by \citet{white2006virtual}. It has influenced various fields, including data visualization and mixed reality, leading to the development of related concepts such as situated analytics~\cite{elsayed2015situated, elsayed2015using}, embedded data representations~\cite{willett2016embedded}, proxsituated visualization~\cite{satriadi2023proxsituated}, and immersive analytics~\cite{ens2021grand, thomas2018situated}, among others. The extent of this work is evident in existing literature reviews~\cite{bressa2021s, saffo2023unraveling, shin2023reality,lee2023design, satriadi2022augmented} and book chapters~\cite{thomas2018situated,schmalstieg2016augmented}. In early work on situated visualization, White~\cite{white2009interaction} highlighted the novelty of this approach, emphasizing how data can be understood through the presence of relevant ``context''. Subsequently, \citet{willett2016embedded} introduced the term ``physical referent,'' which can be seen as an alternative to the term context.

Overall, the idea of situated visualization is to present data using some data representation in the proximity of a relevant context. 
The physical referent is the context of the presented data. 
The contextual information provided by a physical referent has been argued to be valuable in showing the context of a data source and its surroundings.  
Nevertheless, there has been limited investigation into how this context affects cognitive processes such as memory.

Studies on memory in HCI have primarily focused on spatial memory for navigation in virtual environments (e.g., research on memory palace~\cite{krokos2019virtual}) and information seeking in virtual workspaces (e.g., research by \citet{liu2022effects}). In contrast, the context in situated visualization, while partially connected to spatial navigation, predominantly pertains to episodic memory—the recollection of the experience involving the data representation and its surrounding context.

A study worth mentioning in this context is that of \citet{tan2001infocockpit}. In 2001, they introduced the ``Infocockpit'' technique, which involved displaying a panoramic image on a large screen behind standard monitors to enhance the memorability of information presented on the desktop. Their research found that incorporating this panoramic image significantly improved word recall compared to using traditional desktop screens alone~\cite{tan2001infocockpit}.

Nonetheless, there is a limited exploration of similar memory-enhancing experiences. Memorability and context have been studied in the context of password security but yielded no significant effects~\cite{cain2019graphical}.
Furthermore, ongoing research in the field focuses on computer-aided memory using extended reality (XR) technology and wearables. Recently, researchers have investigated the role of XR in manipulating memory. For instance, \citet{bonnail2023memory} presented speculative designs demonstrating how XR can potentially influence personal experiences and knowledge. Similarly, \citet{urakami2023augmenting} introduced a wearable device designed to enhance auditory memory.

\subsection{Summary}
While there is a growing interest in leveraging immersive technology to investigate CDM in learning, the current body of research has largely overlooked its application within the field of data visualization. We observe both shared elements (such as the fundamental concepts of context and target information) and distinctions (including variations in the representation of the target information and the scope of context that extends beyond the immediate environment) when relating recent CDM studies to the realm of situated visualization. However, there is a lack of a clear and organized framework or approach in this area, leaving a gap in the existing literature, which this paper aims to address.

\section{Relevant Design Factors}
\label{designspace}
In this section, we explore the factors to consider when analyzing situated visualization from a CDM perspective. While we acknowledge the presence of several design spaces in situated visualization~\cite{lee2023design, satriadi2023proxsituated, assor2023handling}, it's worth noting that none of them have specifically centered on the study of memorability. We discuss the key factors that should be taken into account when examining situated visualization from the perspective of CDM and clarify their importance.

\subsection{Context-Target Mapping}
In CDM studies, target information is often represented as a list of words, while the environment serves as the context. When environmental features enhance the recall of the target information, these features can be considered contextual memory cues. Conversely, in the opposite scenario, when information is provided but the environmental context must be remembered, it is referred to as source monitoring.

\begin{figure}[htb!]
    \centering
    \includegraphics[width=\linewidth]{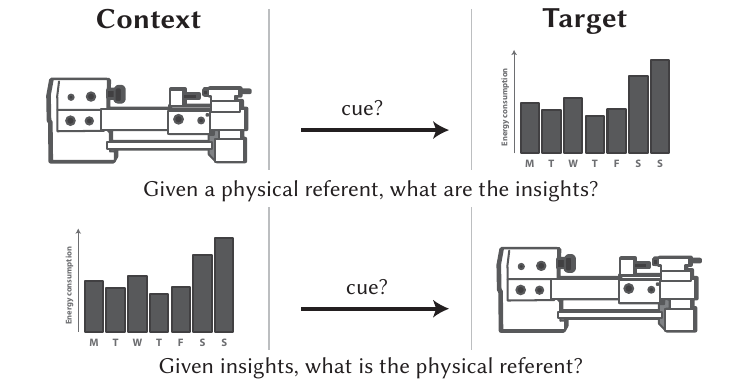}
    \caption{Two possible mappings between context-target and referent-data representation.}
   \label{fig:mapping}
   \Description{The left part of the figure displays two types of context (physical referent and chart), while the right part presents two types of targets (chart and physical referent). The left and right parts are connected by arrows.}
\end{figure}

In situated visualization, the physical referent, serving as target information, can be just as crucial as the data representation itself. As depicted in Fig.~\ref{fig:mapping}, the context-target mapping between the physical referent and data representation is interchangeable. In the top figure, the target information to be recalled is the chart, while the context is the referent. An example question is, "What is the trend of energy consumption for machine $x$?" In the bottom figure, the situation is reversed, with the target information being the referent and the context being the data representation. An example question might be, "Which machine consistently used a large amount of energy during weekends?"

\textbf{Importance}. \textit{Identifying the mapping of context to target is important because remembering information represented as charts might be different from remembering the physical features of physical objects. }

\subsection{Types of Physical Referent}

\begin{figure}[htb!]
    \centering
    \includegraphics[width=1\linewidth]{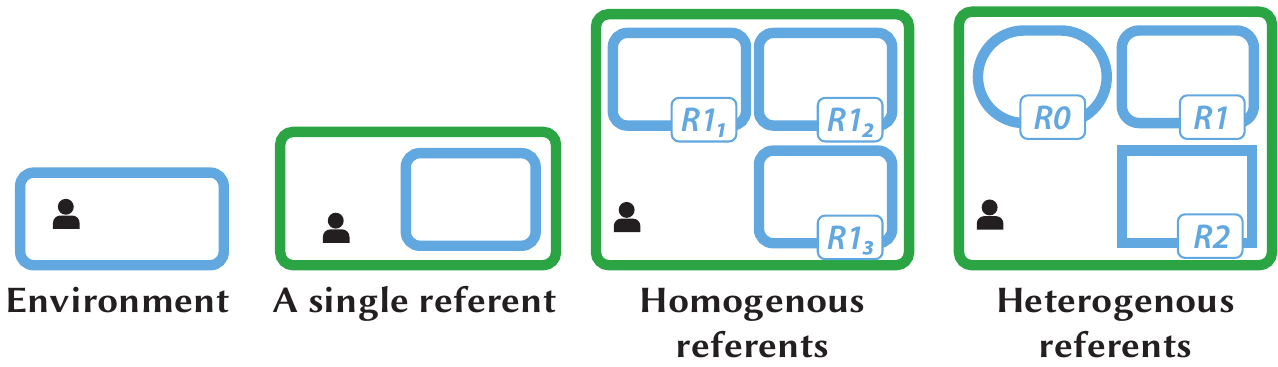}
    \caption{Different types of physical referents (blue shapes). The green shapes represent the environment.}
   \label{fig:type_of_referent}
   \Description{The figure illustrates a conceptual diagram of the environment, a single referent, homogeneous referents, and heterogeneous referents. The diagram employs rounded rectangles to represent different factors.}
\end{figure}

In order to elucidate the complexity of the factors influencing situated visualization and summarize the findings in context-dependent memory research, we must begin by addressing a fundamental question: \textit{What constitutes a physical referent in situated visualization?} In their Ph.D. thesis, White~\cite{white2009interaction} provides a framework for categorizing context, distinguishing it as either a singular referent, multiple referents, or the complete environment (referred to as the `scene' by White). While we adopt this categorization, we further refine the ``multiple referent'' category into homogeneous and heterogeneous referents. Hence, physical referent can be 1) \textbf{environment}, 2) \textbf{a single referent}, 3) \textbf{homogeneous referents}, and 4) \textbf{heterogeneous referents}, as illustrated in Figure \ref{fig:type_of_referent}. 
\begin{enumerate}
    \item \textbf{Environment} denotes a physical referent category that encompasses a spatial domain, often characterized by the interconnections of the objects within it. To illustrate this concept, consider the context of urban data visualization, where a user is positioned on a city street, examining air quality data, as exemplified in the scenario presented in SiteLens~\cite{white2009sitelens}. In such a scenario, the user's perception of the environment serves as the contextual backdrop for the provided information. This encompassing environment includes elements such as the vehicles on the road, the surrounding buildings, pedestrians, and other pertinent environmental factors.
    \item In the case of \textbf{a single referent}, the data primarily pertains to an individual object, with the surrounding environment serving as the context for that referent. 
    \item  \textbf{Homogenous referents}  are a collection of physical entities with minimal distinctive features among them. Examples include a group of zebras or multiple laundry machines. When lacking specific spatial cues, discerning one individual from another within this set of referents can be challenging.
    \item  In contrast, \textbf{heterogeneous referents} consist of individuals with noticeable distinctive features. Examples include students in a classroom, various types of machinery, or buildings with unique architectural designs. 
\end{enumerate}

\textbf{Importance}. \textit{The type of referent plays a crucial role in determining the scope and intricacy of what constitutes context. While an entire environment may encompass a complex interplay of various elements, detailed scrutiny of specific features may not always be necessary. In the case of a single referent serving as the context, one's limited attention capacity can be directed toward the thorough examination of that specific referent's details. Conversely, when dealing with multiple referents, the ability to distinguish each individual becomes paramount. Homogeneous referents, characterized by similar features, may introduce ambiguity in establishing clear associations between each referent and the target information within the referent set. In contrast, referents with heterogeneous features have the potential to enhance memory binding.} 

\subsection{Types of Experiential Situatedness}
\begin{figure}[htb!]
    \centering
    \includegraphics[width=\linewidth]{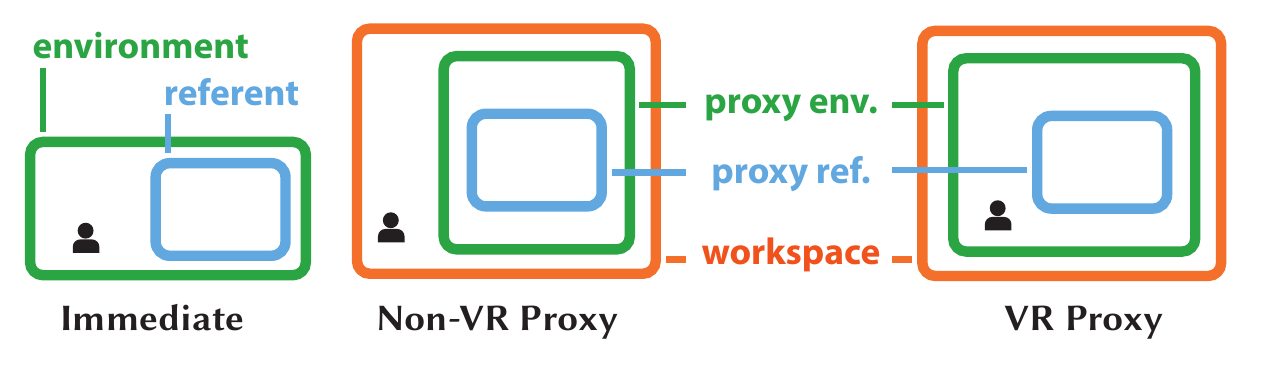}
    \caption{Different types of situatedness define what factors surround the target information and the user.}
   \label{fig:types_of_situatedness}
   \Description{The figure portrays a conceptual diagram of the types of situatedness, including immediate, Non-VR proxy, and VR proxy. The diagram uses rounded rectangles to depict different factors.}
\end{figure}

The type of user experience defines the manner in which context is perceived by the user. In this context, we adopt a recent model of situated visualization, examining the distinction of experiences between immediate situatedness and \textit{proxsituatedness}. \textbf{Immediate} Situated Visualization situates the user and the physical referent within the real-world physical environment. Proxsituated Visualization employs proxy representations of physical referents instead of the actual, immediate physical objects. We consider Non-VR and VR proxsituated visualizations. Figure \ref{fig:types_of_situatedness} illustrates these distinctions. 

In \textbf{Non-VR ProxSituated} Visualization, the user can clearly perceive the workspace around them. This includes, but is not limited to, scenarios where digital models are displayed on desktop screens, physical scale models, or one-to-one digital models presented in augmented reality~\cite{pooryousef2023working}. 

\textbf{VR ProxSituated} Visualization immerses the user in a simulated environment that includes proxy physical referents. In this scenario, the user is physically constrained by the limitations of their physical workspace and may still be aware that they are not actually in the immersive proxy environment. What they perceive during the visualization, however, could be entirely different from their actual surroundings. 

\textbf{Importance}. \textit{
This factor holds significance as it delineates the boundaries of what can be considered as contextual features. For instance, in Immediate Situated Visualization, the context is solely confined to the actual physical environment. However, in both ProxSituated Visualization scenarios, an intricate interplay emerges between the attributes of the user's workspace (such as sound, temperature, and the sense of presence) and the proxy referent and proxy environment. When this factor is combined with the learning and recall phases, it generates a diverse array of scenarios. For instance, one may encounter scenarios involving learning within VR Proxsituated Environments and recalling information within Immediate Environments.
}

\subsection{Data Representation}
\begin{figure}[htb!]
    \centering
    \includegraphics[width=\linewidth]{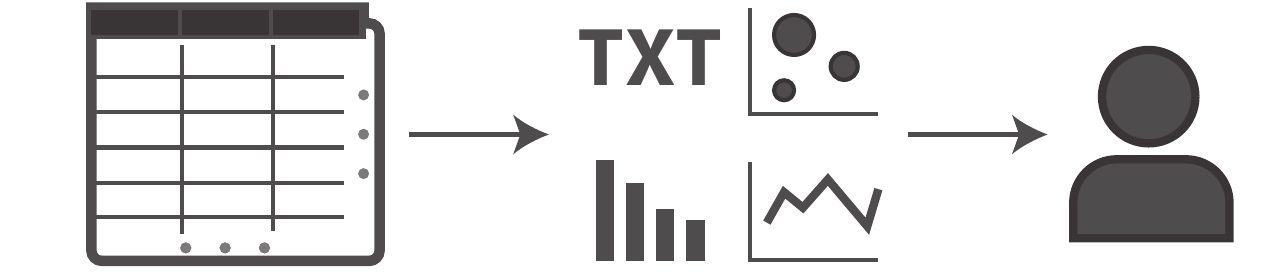}
    \caption{The data representation can have various forms, which might have different effects on memorability.}
   \label{fig:type_of_datarep}
   \Description{The figure illustrates a conceptual diagram of the types of data representation. On the left is a table. In the middle are various data representations, such as text and visualization idioms. On the right is the user.}
\end{figure}

Situated visualization is a method that presents data representations within their relevant context. The complexity and degree of integration of information related to the physical referent can exhibit substantial variation~\cite{satriadi2023proxsituated}. The representation can manifest in various forms (Figure \ref{fig:type_of_datarep}), with the conventional chart being a common example.

Simpler data patterns may employ techniques like color encoding of physical referents or the inclusion of embedded visual elements such as bars or circles. On the other hand, more intricate data patterns may incorporate embedded charts or comprehensive dashboards. Text-based representations are also a viable option, provided that they effectively convey insights derived from the data.

\textbf{Importance}. \textit{ When data representation becomes the focus of recall, it becomes crucial to evaluate its complexity. Key questions arise, such as: How many attributes are encoded within the representation? What is the manner in which these representations are conveyed? It is evident that dealing with complex insights stemming from a multitude of attributes can pose a greater challenge for memory recall. Moreover, the complexity of representations that necessitate interpretation, such as chart reading skills, inherently differs from information presented in a textual format.
}

\subsection{Other factors}
Some other factors that we would like to highlight. 
\subsubsection{Associations}
In cases where the type of data representation can be detached from the referent, such as charts, dashboards, or text panels, the link between information and referent holds significant importance. This connection can manifest through factors like spatial proximity, perceptual similarity, or even semantic relevance, as elucidated in prior research within the field of situated visualization~\cite{thomas2018situated, ellenberg2023spatiality, lee2023design, ellenberg2023spatiality}.
For instance, text that explicitly mentions the referent's name or describes its characteristics establishes a stronger semantic association compared to text that lacks relevance to the referent.

\subsubsection{Retention} Retention refers to the duration between the learning and recall stages. CDM studies have indicated that the impact of contextual cues tends to strengthen with the passage of time. This is grounded in the understanding that as memories naturally undergo a process of drift and decay over time, as observed in research~\cite{smith1994theoretical}, the significance of cues becomes more pronounced precisely when the memory of the target information begins to fade.

\subsubsection {Learning Order}
When dealing with multiple objects, the learning process can take place in either parallel or sequential fashion. Sequential learning, resembling the structured approach commonly found in training activities, facilitates organized learning steps, ensuring that each referent receives a relatively equal share of attention. In contrast, parallel learning, characteristic of exploratory visualization, may lead to unevenly distributed attention, with some referents potentially being overlooked.

\begin{figure*}[htb!]
    \centering
    \includegraphics[width=\linewidth]{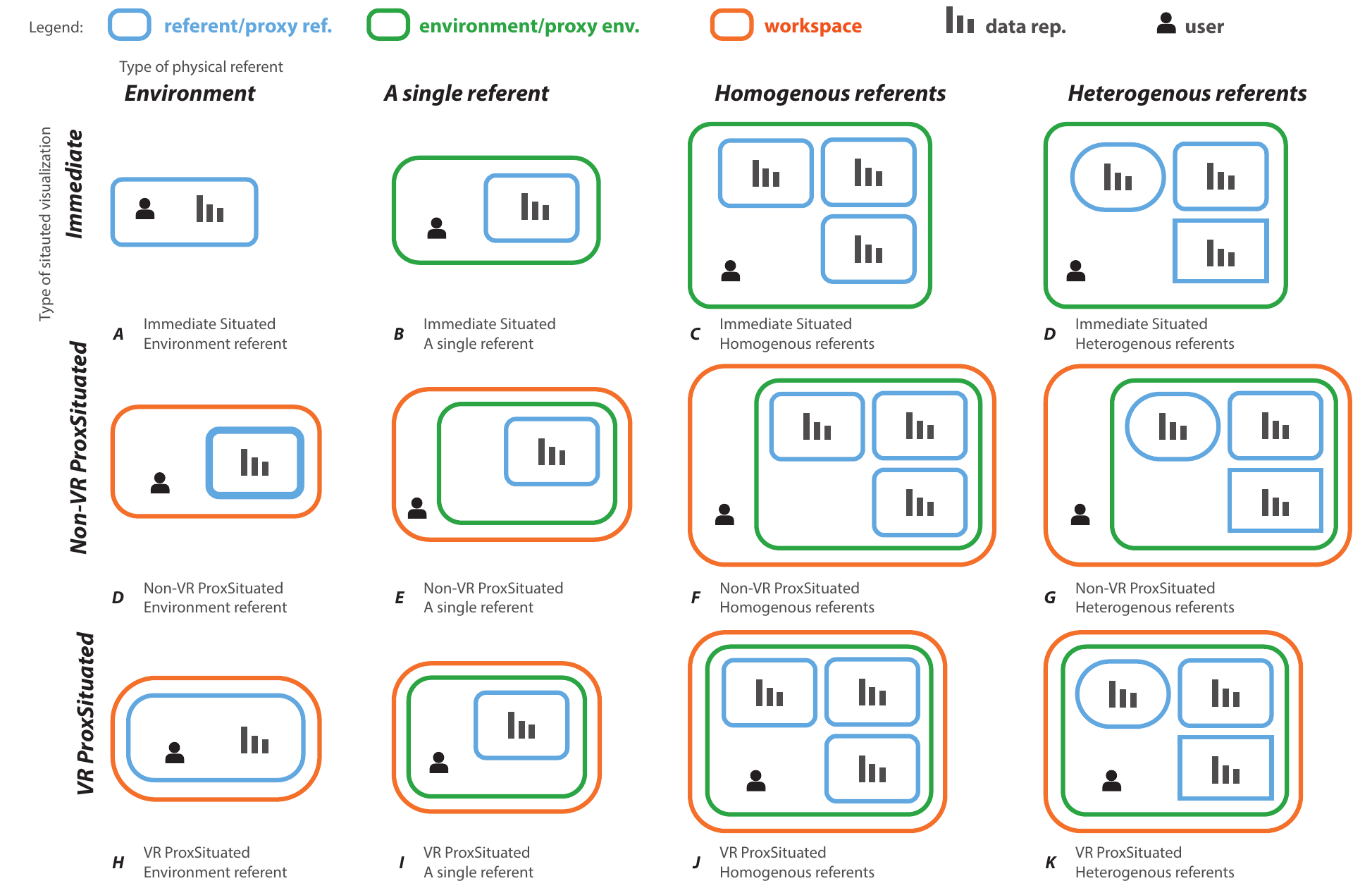}
    \caption{Design space of situated visualization for context-dependent memory study studies, consisting of types of situated visualization (rows) and types of physical referents (columns). The design space illustrates where the user is located, which defines the perceived context during visualization tasks.
    }
   \label{fig:design_space}
   \Description{This is a large figure displaying twelve categories in the design space. The column displays different types of physical referents, such as environment, a single referent, homogeneous referents, and heterogeneous referents. The row showcases different types of situated visualization, including immediate, Non-VR proxy-situated, and VR proxy-situated.}
\end{figure*}

\begin{figure*}[htb!]
    \centering
    \includegraphics[width=\linewidth]{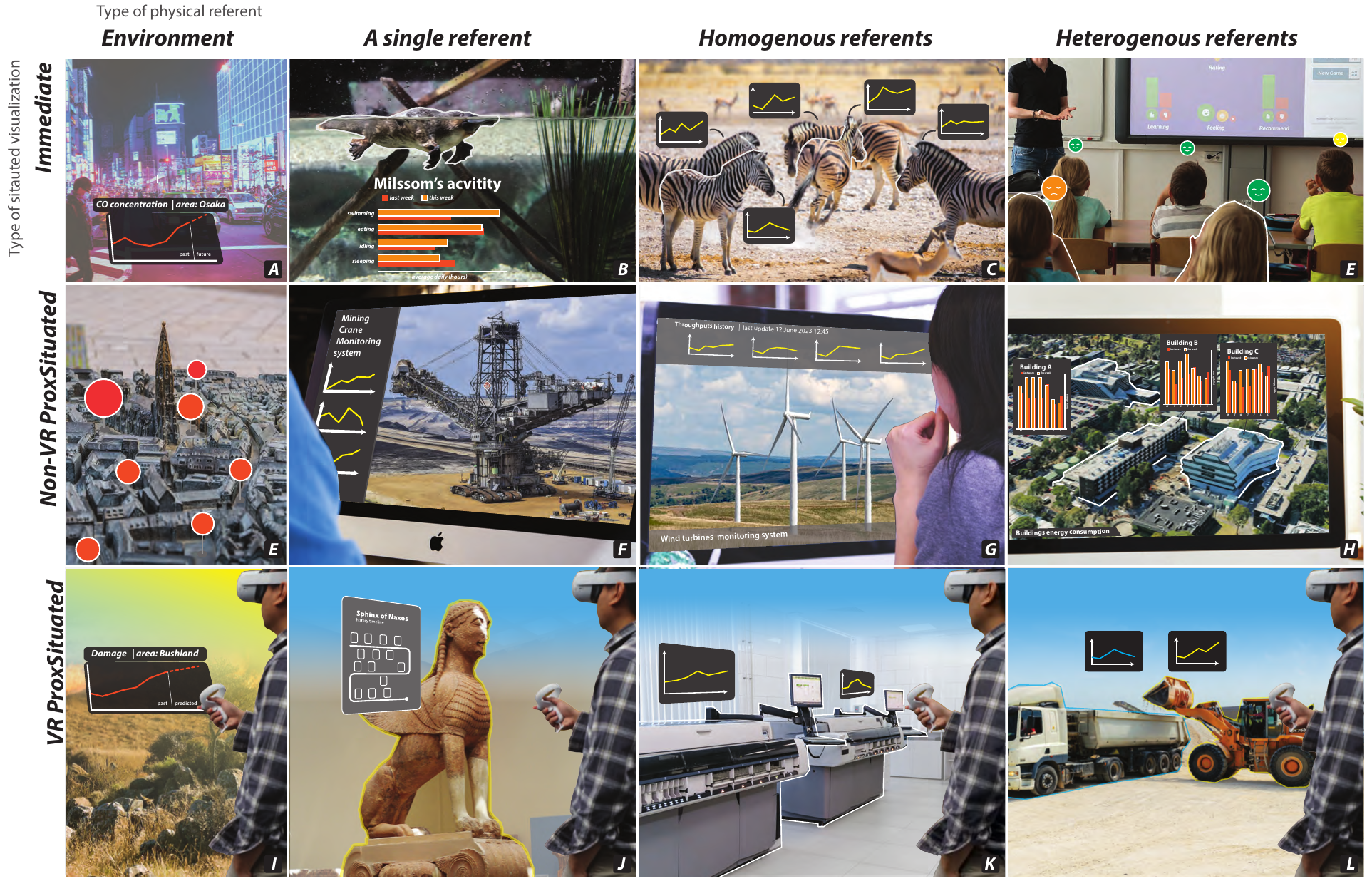}
    \caption{Illustrations of each combination. Top left to bottom right: City environment, platypus, zebras, students, scale model of a city, a video feed of a mining crane, a video feed of wind turbines, models of campus buildings, bush fire environment, Sphinx of Naxos, laboratory equipment, mining vehicles. }
   \label{fig:examples}
\end{figure*}

\subsection{Design Space}
The combination of the types of physical referents and the types of situatedness experience factors creates a design space for understanding CDM in the situated visualization domain. While there are interactions with other factors, we consider the type of physical referent and type of situatedness to be crucial. The generated design space contains twelve possible situated visualization scenarios, as illustrated in Figures \ref{fig:design_space}. The former depicts the concepts, while the latter figure portrays realistic scenarios. The interaction between these two factors results in different combinations of factors that we can consider as context.

We can illustrate these concepts with a spectrum of examples. The simplest one involves immediately situated visualizations, where the entire environment serves as the context (denoted as "A" in Figures \ref{fig:design_space} and \ref{fig:examples}). In contrast, a more intricate scenario presents itself in Non-VR proxsituated visualizations with heterogeneous referents, represented as "H". In this case, users engage with their physical workspace while simultaneously focusing their attention on the proxy environment, proxy referent, and the data representation.
Likewise, the VR version of the same scenario (designated as "L") immerses the user within a simulated proxy environment. However, even in this immersive context, they may retain awareness of their surrounding workspace through non-visual senses like vestibular cues and their preexisting knowledge of the physical space. For a concise summary of each category, please refer to the Appendix \ref{appendixdesignspace}.

\section{Studies: Heterogeneous Referents in ProxSituated Scenarios} 
\label{studies}
\subsection{Motivation}
As demonstrated in the previous section, the domain of CDM and situated visualization is extensive, yet largely unexplored. This paper aims to initiate exploration within this domain, although selecting a suitable starting point is not straightforward. We have opted for a pragmatic approach, bypassing the simplest scenarios. Our research is motivated by prox-situated scenarios~\cite{satriadi2023proxsituated}, which involve the use of physical referent proxies. In certain cases of data visualization and visual analytics, the inclusion of these referents often introduces additional complexities, such as modeling and rendering. Alternatively, visual analysis can be conducted without situational context, focusing solely on data representation. Our objective with these studies is to gain a better understanding of the contextual value of such proxies, particularly in terms of information memorability. We concentrate on heterogeneous referents presented through proxies on both desktop and immersive VR platforms, as these scenarios hold potential applications across various domains.

In particular, our research question is ``\textbf{{If the learning and recall environments are the same, can the presence and absence of proxy physical referent alone affect one's ability to remember data representation?}}'' The answer to that question will provide a baseline for our understanding of the contextual cues provided by the physical referent proxy.  
\begin{figure*}[htb!]
    \centering
    \includegraphics[width=0.7\linewidth]{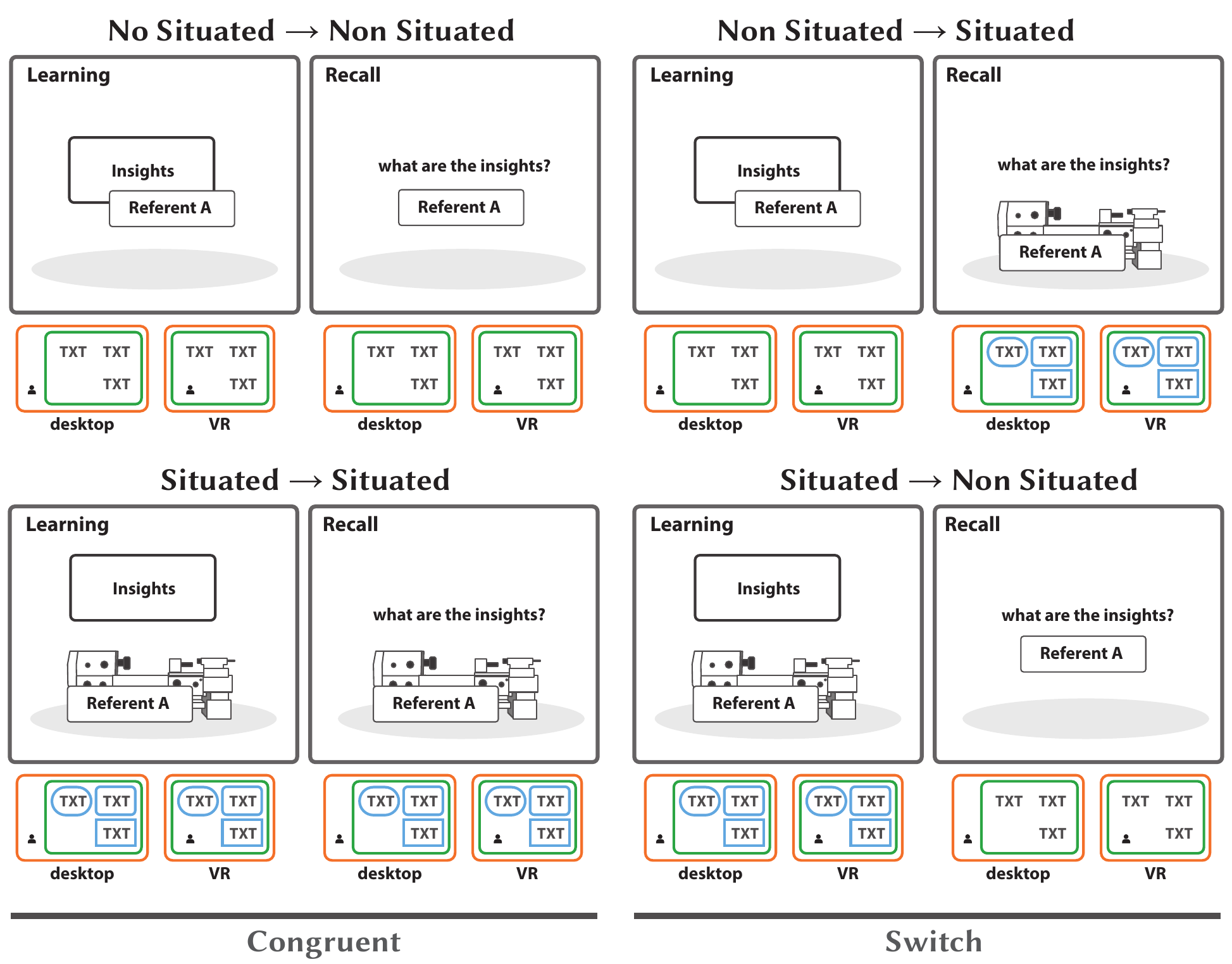}
    \caption{Four between-subject \defgroup{} conditions evaluated in our study are \defsitusitu{}, \defsituNONsitu{}, \defNONsituNONsitu{}, \defNONsitusitu{}. From these groups, there are two types of congruency. Congruent groups have the same learning and recall environments (\defNONsituNONsitu{}, \defsitusitu{}), while the Switch groups have different environments (\defNONsitusitu{}, \defsituNONsitu{}).}
   \label{fig:study_design}
   \Description{The figure displays illustrations of the four groups tested in the studies. Beneath each group are conceptual diagrams of the desktop and VR conditions.}
\end{figure*}

\subsection{Study Factors}
Our experiments consist of multiple studies involving two main factors: \textit{modality} and \textit{group}, as illustrated in Figure \ref{fig:study_design}.

\subsubsection{Studies in two modilities}
The first factor is the type of experiential situatedness, from which we choose desktop and virtual reality modalities. We ran three studies in total, two studies on \textit{desktop} and one in \textit{VR}. The reason for the two desktop studies is discussed later. 

\subsubsection{Within Group}
Within each study, there were four groups that are the result of the combination between Phase (\defLearning{} and \defRecall{}) and Situatedness (\defSitu{} and \defNonSitu{}) factors, as illustrated in Figure \ref{fig:study_design}. For example, situ learning and non-situ recall mean that the participants were only seeing the proxy referent during the learning but not during the recall.
Therefore, the four groups are: \defsitusitu{}, \defsituNONsitu{}, \defNONsituNONsitu{}, \defNONsitusitu{}. We use a between-subject design for these groups. Note that the term ``Situ'' we use from this point refers to  proxsituated visualization where data representations are associated with physical referent proxy, rather than immediate in-situ visualization.  

\subsubsection{Within Group Referents}
Within each group in each study,  we use three types of heterogenous referents varying in size and physical form. We counter-balance the order their presentation per group.

\subsubsection{Within Group Target Information}
The information is a sentence consisting of two pairs of attribute $+$ pattern. 
We picked the pairs randomly from the list of attributes and patterns but made sure that there were no repeated pairs across referents. Within each referent, the attributes cannot be repeated while repeated patterns are allowed.  

\begin{figure}[h!]
    \centering
    \includegraphics[width=\linewidth]{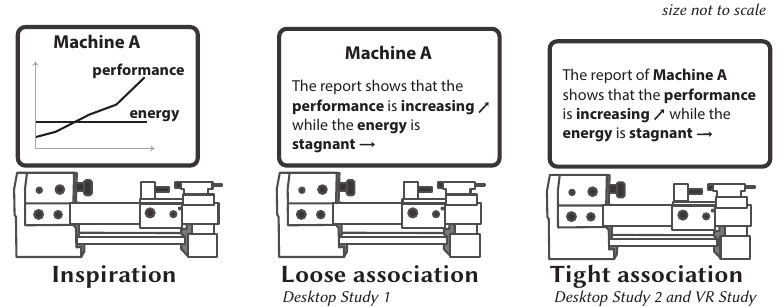}
    \caption{Insights presented in charts (left) inspired the information in our study (right) . This approach is to avoid co-founding factors from varying chart reading abilities of the participants.}
   \label{fig:conversion}
   \Description{The left part of the figure illustrates a machine with a chart view displaying a line chart. The middle and right figures depict similar illustrations that show text instead of charts in the views.}
\end{figure}

\subsection{Learning Task}
\subsubsection{Mode of Learning}
Mode of learning can be categorized as intentional and incidental (see ~\cite{plancher2010age, smith1994theoretical}). 
We focus on intentional learning as most current scenarios for situated visualization (at least in industry) are for data analysis, i.e. learning about the data, as opposed to communicative visualization for public. 

\subsubsection{Information Complexity}
In our study, we attempt to simulate a data pattern that is not too complex yet not too simple to memorize. We use two pairs of attribute $+$ pattern pair per referent. For example, a pair can be ``the performance [\textit{attribute}] is increasing [\textit{data pattern}]''. 
Existing work in situated visualization presents data in a simple representation such as in color-coding or embedded marks~\cite{satriadi2023proxsituated}. We focus on scenarios with more complex representations such as embedded charts or embedded dashboards (e.g.,~\cite{satriadi2022augmented, alonso2018cityscope, prouzeau2020corsican,whitlock2020hydrogenar, lin2022quest, cunningham2021towards, ellenberg2023spatiality}). 

\subsubsection{Target Information}
Related work in the CDM domain has used low-level tasks such as memorizing words or sentences. In situated visualization, the information will be insights presented in the chart. These insights, however, are highly dependent on the chart-reading capability of the user. To reduce possible confounding factors caused by discrepancy in chart-reading skills, we used text that is enhanced with simple icons for the data patterns (e.g., up-arrow for increasing, down-arrow for decreasing, horizontal arrow for stagnant), as shown in Figure \ref{fig:conversion}-middle and right. 

We also looked at two types information presentation that vary in their association between the referent and target information. In data visualization, it is common to label the chart with the name of the referent, e.g., Machine A (as shown in Figure \ref{fig:conversion}-left). Converting this representation directly into text would result in the type of presentation shown in Figure \ref{fig:conversion}-middle where the text box has separated referent label and insights. We refer to this as \textit{loose association} as the insights do not contain any direct association with the referent, i.e.\ the participants might only memorize the text insights and ignore the label. A stronger type of association would be the one shown in Figure \ref{fig:conversion}-right where the label of the referent is integrated with the target information. We refer to this as \textit{tight association}. Note that we only use the terms loose and tight for referencing purposes in the paper, rather than claiming significant differences or correllations between the two.  

Two of our studies on desktop modality investigate the loose and tight association conditions while the VR study was only conducted with the tight association. This is due to logistics as collecting data for VR studies is costly and tight association will ensure that the label of the referent is learned by the participants.

For each referent, there are pairs of data attributes and patterns. An example of a pair would be ``\textit{\textbf{breakdown} is \textbf{increasing}}''. For each data attribute, there are four options. For each data pattern, there are six patterns.

\subsubsection{Referent Proxy}
The complexity of referent is difficult to quantify and can be subjective. Just recently, a database of objects for psychology studies was published~\cite{popic2020database}. This database contains various everyday objects and their subjectively measured complexity. 
Using this database would allow us to investigate the effect of referent complexity and information memorability. However, the use of familiar objects might lead to decontextualization~\cite{smith1994theoretical}, i.e., the loss of connection between the environment and information because the information or environment is too common. Moreover, we focus on situations where context from non-every-day objects matters. Limiting the type of referent to an unfamiliar object is a fair way to reduce subjective bias. We decided to use models of industrial machines, which we assume, are not common for most people and fit well with in-the-wild applications. 

\begin{figure}[htb!]
    \centering
    \includegraphics[width=\linewidth]{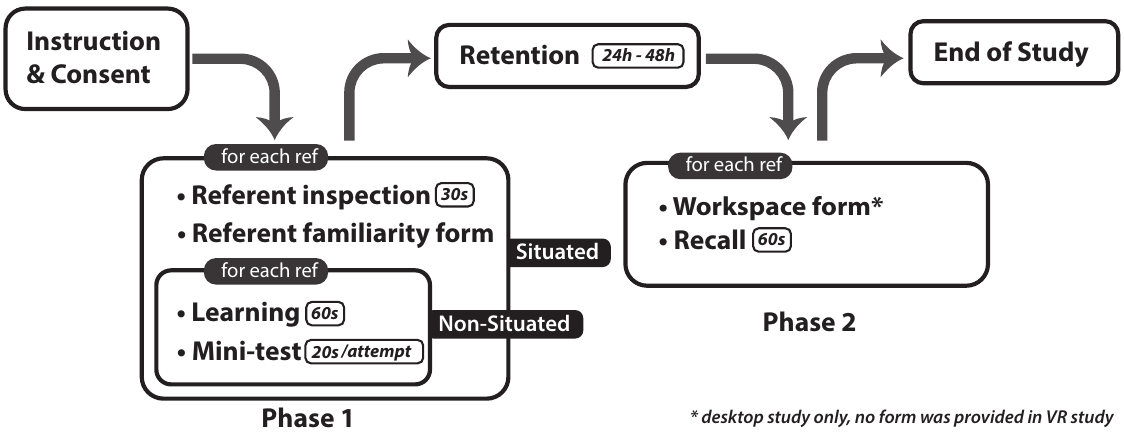}
    \caption{General flow of the studies. The ``for each ref'' represents steps within each presented proxy physical referent, i.e. machine.}
   \label{fig:procedure}
   \Description{The figure shows the flow of the study. There are four main parts represented as rectangles: introduction, learning phase, retention, recall phase, and end of study.}
\end{figure}

\subsection{General Procedure}

The main procedure consists of the learning of information and a mini-test for each referent presented in a Latin square order. The overall steps are shown in Figure \ref{fig:procedure}. A few details are as follows. 

\begin{figure*}[ht!]
    \centering
    \includegraphics[width=\linewidth]{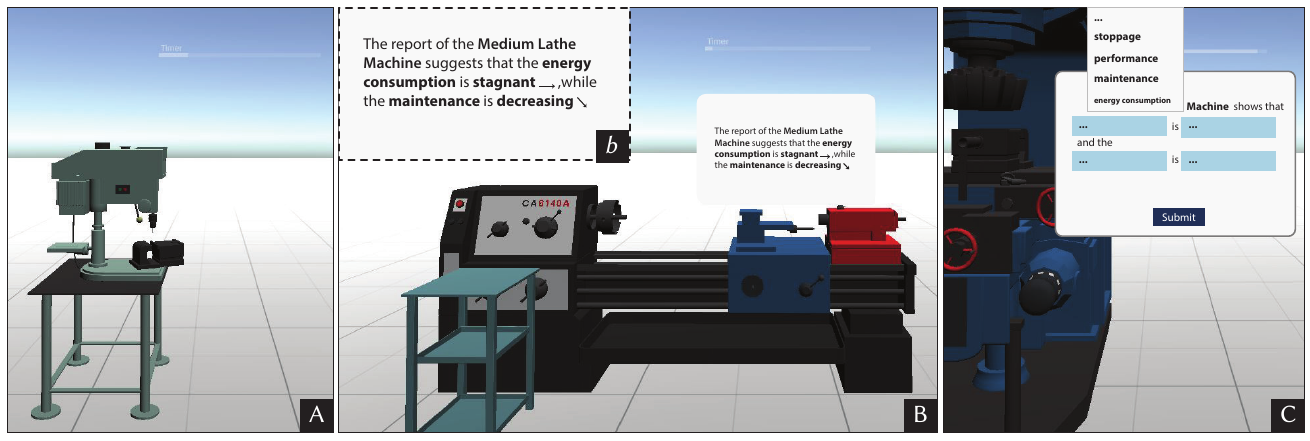}
    \caption{Left figure (A) is the machine introduction stage where the participant is shown the machine and asked to rate their familiarity. The middle figure (B) is the learning stage where the target information is presented. The inset (b) is the zoomed in of the target information. In non-situ conditions, only the information panel is presented. The right figure (C) shows the answer panel in the mini-tests. These three model of the machines are used in our study.}
   \label{fig:study_setup_vr}
   \Description{The left figure shows an empty VR scene with a 3D model of a small machine. The middle figure displays an empty VR scene with a 3D model of a medium-sized machine and a text view attached to it. The right figure exhibits an empty VR scene with a 3D model of a large machine with a fill-the-blank form attached to it.}
\end{figure*}

\begin{enumerate}
    \item 
        \textbf{Consent and instruction}
A high-level description of the research and procedure was provided to participants prior to commencement. 
\item 
\textbf{Induction (only in Situated Learning)}
For the situated learning group, there is a 30 seconds phase where we allow participants to inspect the reference before any target information is displayed (Figure \ref{fig:study_setup_vr}-A). 
After inspecting the referent, we asked them to fill in a questionnaire about their familiarity with the presented machine.

In the desktop study, we rotate the model along the vertical axis to show the 3D shape. We did not rotate the object in VR study as the stereoscopic view was enough to create 3D perception. 

\item
\textbf{Learning information}. We showed target information, i.e. the two pairs or attribute $+$ pattern (Figure \ref{fig:study_setup_vr}-B), for 60 seconds per referent during the learning phase.
\item
\textbf{Mini test}. After learning for each referent, we asked the participants to answer `fill-in-the-blanks' questions where we hid the attributes and patterns. 
The participants selected the most relevant attribute or pattern using drop-down menus. This means they selected one answer from a list of options. There were 4 drop-down menus in total (2 pairs). The participants submitted their answers by pressing a submit button (see Figure \ref{fig:study_setup_vr}-C).
We gave them 3 chances per mini-test. Each attempt was limited to 20 seconds. 
If their answers were correct, they will be redirected to the next step. 
Otherwise, they needed to try again until the answers were correct or they ran out of attempts. In both cases, the participant will see the answer for another 10 seconds before being taken to the next step. 

\item 
\textbf{Recall workspace questionnaire (only in Desktop studies)}.
Before starting the recall session, we asked the participants to describe the difference between their learning and recall workspace. 
\item 
\textbf{Recall test}. The order of the referents in the recall phase was different from the learning phase. The recall phase order is the next item of the learning phaser order in the Latin square table. 
The time limit for each referent was 60 seconds. 
The interface is the same as the mini-test interface.
We made clear in the instruction that they only had one attempt. 
\item
\textbf{Retention}. Previous studies suggest that CDM effects are more apparent in long-term memory than short-term. We set the retention time to be between 24 and 48 hours after the completion of the learning phase. This is expected to allow for long term recall and a sufficiently wide time window for allow participants enough flexibility to come back.
\item 
\textbf{Timer indicator}. 
A timer bar was shown on each timed step. 
\end{enumerate}

\subsection{Technical Details}
Our desktop study instrument is a web application built using the combination of Slim PHP Framework and Twig as the backend.
We used Three.js to render the models in situated conditions. 
We also built a simple API using the same web application. 
This API was used by the VR study application.
The main application of the VR study was built using Unity and MRTK. 
We built it into WebXR-supported WebGL application using the SimpleWebXR\footnote{\url{https://github.com/Rufus31415/Simple-WebXR-Unity}} repository.
We served the web application (for desktop) and the WebGL application (for VR) on our server.

We used a single link entry for both phases and determined the participant's status in our system by recording their designated participant ID. Upon completion of Phase 1, we showed the instruction to come back for Phase 2. \defVR{} participants were allowed to use Windows Mixed Reality headsets or Meta Quests headsets.

\subsection{Data Collection}

\subsubsection{Participants}
The participants of our desktop studies were gathered through a crowd-sourcing service named Prolific.co. We collected 48 participants per study (12 participants x 4 \defgroup). We set a constraint in the platform so that all 96 participants for the two studies are unique (none of the participants participated in both studies). For the VR study, we used convenient sampling by inviting people from our professional network as well as placing posters at our university. Eight (8) participants participated in their location (our study setup is accessible through a web browser), while the remaining 40 participants performed the study in our lab.

\subsubsection{Compensation}
For both modalities, we made clear in the instruction that the participants were only compensated if they completed both phases. 
We paid the crowdsourcing participants (desktop studies) \textsterling 2.25  which is within \textsterling 9/hour rate (considered ``good'' by the Prolific system), for the desktop study. 
For the VR study, we provided the participants with a \$20 gift card for their participation. Both desktop and VR studies last for approximately 15 minutes.

\subsubsection{Measures}
Our primary measure is the accuracy of the recalled information. We also collected other measures including the following. 
\begin{itemize}[leftmargin=*]
\item \textbf{Demographic}. This includes extensive demographic data from Prolific for desktop studies and basic demographic information for VR studies. 
    \item \textbf{Learning accuracy}. This data contains the information of each attempt performed during the mini quizzes. This allows us to see how success the learning was and to see if the order of the referent matters during the learning process. 
    \item \textbf{Environment conditions (only in desktop studies)}. This data contains the information regarding the surrounding of the participant during the learning and recall. We asked the information regarding room, desk, computer, lighting condition, and their posture to which the participants answer either ``same'', ``slightly different'', and ``completely different''. 
    \item \textbf{Referent familiarity (only in situ learning)}. This data is gathered through a form where the participants select either ``Never seen such a machine'', ``I am familiar with such a machine'', or ``I use such a machine regularly''. We focused on the participants familiarity with the shape of the referents, and thus, only gathered this data in the two situe learning \defgroup. 
    \item \textbf{Browser data (only in desktop studies)}. This includes the screen resolution and the size of the browser windows used during the learning and recall phases. 
\end{itemize}

\subsubsection{Demography}
In \defDesktopOne{}, the participants consist of 16 females and 32 males. These are with varying experienced Prolific users with average 327  approvals ($min$ = 28, $max$ = 1793, $SD$ = 329). Average age of the participants is 29.53 ($min$ = 19, $max$ = 64, $SD$ = 9.9). 

In \defDesktopTwo{}, the participants consist of of 21 females, 26 males and 1 gender not stated. These are with varying experienced Prolific users with average 701 approvals ($min$ = 51, $max$ = 3712, $SD$ = 806). Average age of the participants is 30.41 ($min$ = 19, $max$ = 55, $SD$ = 10). 

In \defVR{} study, the age grup distribution is as follows: 18-24 (11 participants or 23\%), 25-34 (34 participants or 71\%), and 35-44 (3 participants or 6\%). Ten participants did not provide their occupation, while the rest are Bachelor students, Master students, PhD students, academic staff, and researchers.

\subsection{Measuring Accuracy}
Accuracy of information being recalled has been used to determine one's memory performance and there are multiple ways on how we can measure accuracy. There are two things to consider. First, how should we define correct answers?
Second, how should we calculate accuracy from correct answers? 

There are 3 referents (i.e. machines). For each referent, there are 2 pairs of attribute (a) and pattern (p). One example will be \textbf{the \defattributeone{breakdown (a1)} is \defpatternone{increasing (p1)} and the \defattributetwo{performance (a2)} is \defpatterntwo{stagnant (p2)}}.
Each data attribute has 4 options (stoppage, performance, maintenance, and energy consumption).
Each data pattern has 6 options (increasing, decreasing, fluctuating, stagnant, peaking, plunging).

\subsubsection{How to define correct answers?}
To define whether or not the participant's response is correct, we follow two principles. 
\begin{itemize} [leftmargin=*]
    \item \textbf{The pair matters}. In common memory studies, each word in the word list is self-contained (e.g., car, book, jump, travel, etc.). In visualization and our study, being able to correctly recall the pattern of given attribute is crucial. In the previous example, \defattributeone{breakdown} must be paired with \defpatternone{increasing} and \defattributetwo{performance} must be paired with \defpatterntwo{stagnant}. 
    \item \textbf{The order of the pair does no matter}. The difficulty or complexity of the information is affected by whether or not the order matters. From practical point of view, the order does not really matter. ``\defattributeone{Breakdown}  is \defpatternone{increasing} and \defattributetwo{performance}  is \defpatterntwo{stagnant}'' is the same as ``\defattributetwo{performance} is \defpatterntwo{stagnant} and \defattributeone{breakdown} is \defpatternone{increasing}''.

\end{itemize}

\subsubsection{How to calculate accuracy from correct answers?}
Now that we have figured out how to define correct answers, we need to consider how accuracy should be calculated. 
A naïve approach would simply sum up correct answers per group. The number of responses per participant is 12 (3 referent x 2 attribute x 2 pattern). This means 12 is 100\% correct answer. We use a weighted approach that weights the correct answer proportionally to the number of available options, as described in the following formula.  
\[ score = ( 4a_1 + 6p_1 + 3a_2 + 6p_2) / 19 \]

Attributes and patterns (i.e., $a_1$, $p_1$, $a_2$, $p_2$) are either 1 or 0 (based on the correctness criteria mentioned in the previous section). Each weight represents the number of options. The second correct attribute is given a score of 3 rather than 4 because after choosing the first attribute, there are only 3 possible answers. Each pattern is weighted as 6 as there are six options and these patterns can be repeated. From the accuracy score, we create ordinal groups that range from 0 to 5, as shown in Table \ref{table:calculation}. The algorithm and implementation of the accuracy calculation are provided in the Supplementary Material. 

\begin{table}[]
\caption{The accuracy (acc.) table depicting various levels of recall performance used in our analysis.}
\label{table:calculation}
\footnotesize
  \taburulecolor{lightgray}
\begin{tabu}{l|c|c|c|c|l|c}
\hline
\textbf{response}                      & \multicolumn{1}{l|}{\textbf{a1}} & \multicolumn{1}{l|}{\textbf{p1}} & \multicolumn{1}{l|}{\textbf{a2}} & \multicolumn{1}{l|}{\textbf{p2}} & \multicolumn{1}{l|}{\textbf{score}} & \multicolumn{1}{l}{\textbf{acc.}} \\ \hline
\hline
all pairs are correct                  & $ \checkmark $                                & $ \checkmark $                                 & $ \checkmark $                                 & $ \checkmark $                                 & 1.00                                 & 5                                     \\ \hline
one pair and one attribute are correct & $ \checkmark $                                 & $ \checkmark $                                & $ \checkmark $                               & $ \times $                                 & 0.68                                  & 4                                     \\ \hline
one pair is correct                    & $ \checkmark $                                 & $ \checkmark $                                 & $ \times $                                     & $ \times $                                     & 0.53                                  & 3                                     \\ \hline
two attributes are correct             & $ \checkmark $                                & $ \times $                                    & $ \checkmark $                                 & $ \times $                                     & 0.37                                   & 2                                     \\ \hline
one attribute is correct               & $ \checkmark $                                & $ \times $                                     & $ \times $                                     & $ \times $                                     & 0.21                                   & 1                                     \\ \hline
all other combinations                       & $ - $                                     & $-$                                     & $ - $                                     & $ - $                                     & 0.00                                   & 0                                     \\ \hline
\end{tabu}
\end{table}
 
\subsection{Analysis}
For the main accuracy measure, the response and answer of our study is a set of words consisting of two pairs of attributes and data patterns.
 For each study, we collected 144 responses (12 participants/group $\times$ 4 groups $\times$ 3 referents/participant). We used a combination of visualization, descriptive statistics, and hypothesis testing for our analysis. As described in the previous section, the accuracy data is on an ordinal scale ranging from 0 to 5. We use the Kruskal-Wallis test and One-way ANOVA for this data. We applied Aligned-Rank Transform~\cite{wobbrock2011aligned} for the ANOVA test if the normality assumption of the raw data is not met. For analyzing the effect of the referent order, we used a separate Friedman test for each group and ART Repeated-Measure ANOVA as the order is within-subject per \defgroup{}. In the instance of statistically significant results, we used Wilcoxon Tests with Bonferroni correction. The analysis is provided in the Supplementary Material.

\section{Results}
\label{results}
The results are presented by the measure category, rather than per study. This will allow us to see characteristics of different \defModality{} and \defgroup{} without repetitive structure. We refer to the studies as \defDesktopOne{}, \defDesktopTwo{}, and \defVR{}. 

\subsection{Retention}
\begin{figure}[htb!]
    \centering
    \includegraphics[width=\linewidth]{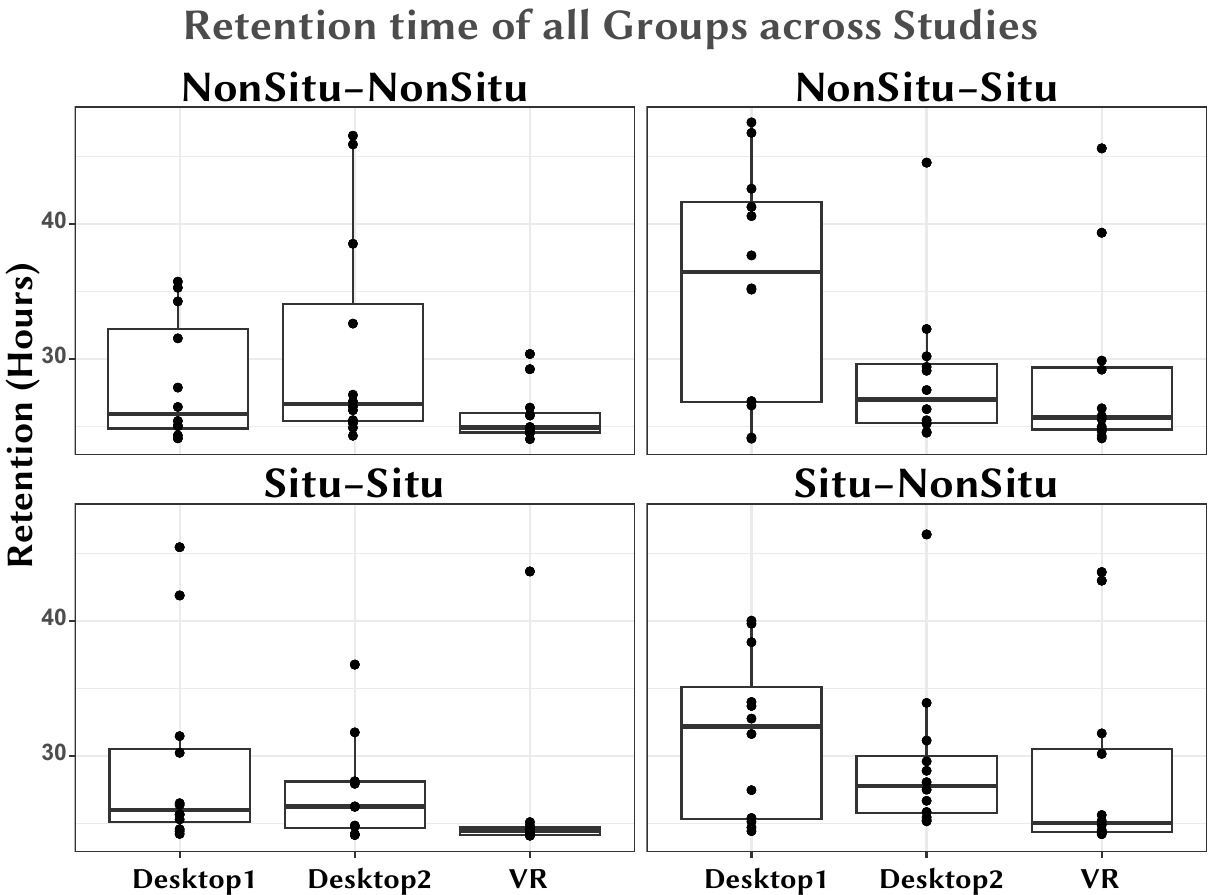}
    \caption{Distribution of memory retention duration of all \defgroup across different \defModality.}   \label{fig:plot_retention_group}
    \Description{The figure displays a small multiple of boxplots showing retention hours of the three studies in four conditions.}
\end{figure}

The retention of the entire study is averaging between 27 to 31 hours. 
The average retention of study \defDesktopOne{} is 31.2 hours ($SD$ = 7.1 hours). The average retention of study \defDesktopTwo{} is 29.1 hours ($SD$ = 6 hours). The average retention of \defVR{} study is 27.4 hours ($SD$ = 5.8 hours). 
Except for \defNONsitusitu{}{} and \defsituNONsitu{} in \defDesktopOne{} that have retention above 30 hours, most groups across studies have retention between 24 to 30 hours (Figure \ref{fig:plot_retention_group}). 

\subsection{Learning Performance}
Learning performance reflects how well the participants performed during the mini quiz, which can be used as a proxy to how successful the learning was. 

\subsubsection{\defDesktopOne{}} Most responses during the learning phase mini quiz were performed with a single attempt (73.1\%, 144 out of 197), while 19.3\% were performed in 2 attempts, and 7.1\% completed in 3. There is one response with 4 attempts, which was an error in the data input and was supposed to be a single attempt. Overall, 71.1\% of the learning steps were completed in the highest possible accuracy. 

\subsubsection{\defDesktopTwo{}} 65.7\% mini quizzes were completed with a single attempt, 26.0\% with two attempts, 7.7\% with 3 attempts. There is no substantial differences among different groups (mostly completed with high accuracy).

\subsubsection{\defVR{}} 77.83\% mini quizzes were completed with a single attempt, 17.29\% with two attempts, 4.8\% with 3 attempts. There is no substantial differences among different groups (all completed).

\subsection{Accuracy}
The accuracy reflects the completeness of the information recalled by the participants. For this measure, we also looked at the combination across \defModality. 

\begin{figure*}[ht!]
    \centering
    \includegraphics[width=\linewidth]{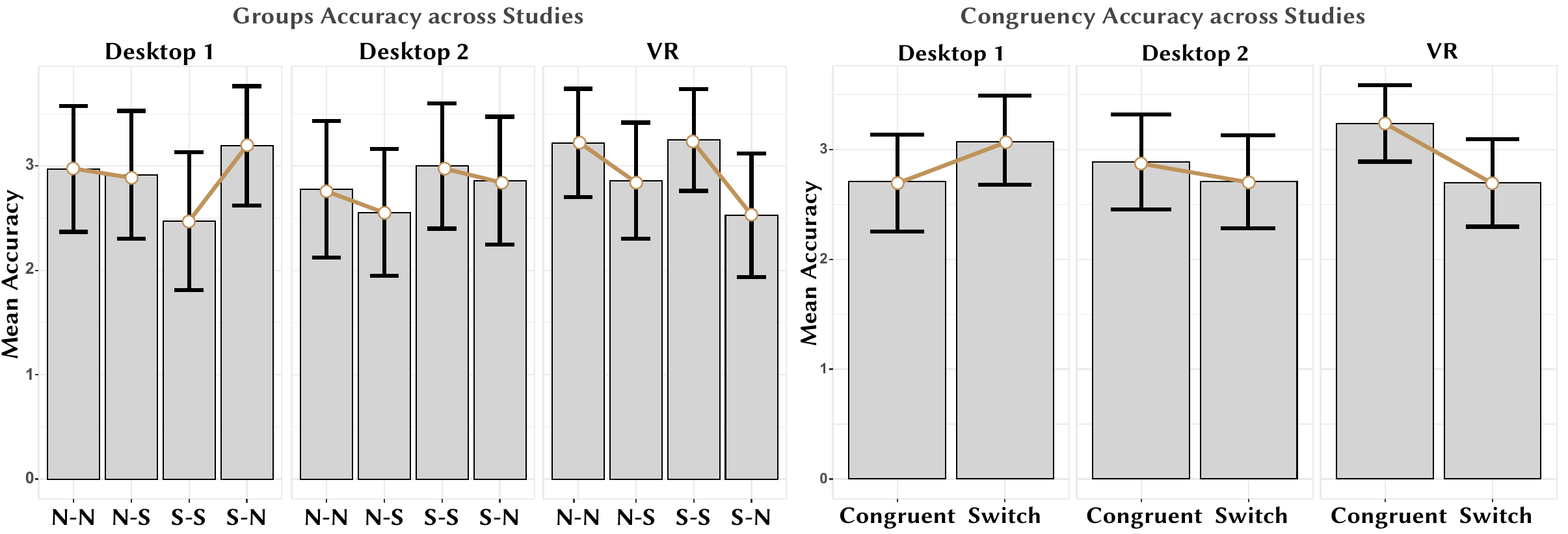}
    \caption{Accuracy per group and congruency in different studies. The error bar represents 95\% CI. The abbreviation N means NonSitu and S means Situ.}   
    \label{fig:plot_accuracy_combined}
    \Description{The figure presents 2 groups of bar charts. In each group, there are 3 bar charts representing the recall accuracy in each study.}
\end{figure*}

\subsubsection{\defDesktopOne{}}
We did not find any significant differences between the four tested conditions (Kruskal-Wallis $\tilde{\chi}^2$ = 2.9437, df = 3, p-value = 0.4004) and between congruency factors (Kruskal-Wallis $\tilde{\chi}^2$ = 1.3565, df = 1, p-value = 0.2441
). With a loosely connected information and referent, i.e. when the label of the referent is not integrated with the text, the presence of the physical referent does not affect the reader’s ability to memorise the information

\subsubsection{\defDesktopTwo{}}
We did not find any differences between the four tested conditions (Kruskal-Wallis $\tilde{\chi}^2$ = 1.2574, df = 3, p = 0.7393) and between congruency factors (Kruskal-Wallis $\tilde{\chi}^2$  = 0.54194, df = 1, p = 0.4616
). With a closely connected information and referent, i.e. when the label of the referent is integrated with the text, the presence of the physical referent does not affect the reader’s ability to memorise the information.

\subsubsection{\defVR{}}
We did not find any evidence of the effect of the group on the accuracy (Kruskal-Wallis $\tilde{\chi}^2$ = 4.3912, df = 3, p = 0.2222). Congruency shows trend but not significant (Kruskal-Wallis $\tilde{\chi}^2$ = 3.7379, df = 1, p = 0.05319). The summary statistics are provided in the Table \ref{table:acc_group_vr} and Table \ref{table:acc_congruency_vr}.

\begin{table}[]
\caption{The summary statistics of the accuracy across the \defgroup{} in \defVR{} study.}
\label{table:acc_group_vr}
\footnotesize
\taburulecolor{lightgray}
\begin{tabu}{l|l|l|l|l|l|l|l}

\hline
\textbf{Modality} & \textbf{Group} & \textbf{mean}     & \textbf{median} & \textbf{sd}       & \textbf{n}  & \textbf{se}        & \textbf{ci}        \\ \hline
\hline
\defDesktopOne{}      & NonSitu-NonSitu & 2.97     & 3.0    & 1.78     & 36 & 0.26      & 0.60      \\ \hline
\defDesktopOne{}       & NonSitu-Situ    & 2.92     & 3.0    & 1.81     & 36 & 0.30      & 0.61      \\ \hline
\defDesktopOne{}       & Situ-Situ       & 2.47     & 2,0    & 1.95     & 36 & 0.32      & 0.66      \\ \hline
\defDesktopOne{}       & Situ-NonSitu    & 3.19     & 4.0    & 1.69    & 36 & 0.28      & 0.57      \\ 
\hline
\hline
\defDesktopTwo{}       & NonSitu-NonSitu & 2.78     & 2.5    & 1.93     & 36 & 0.32      & 0.65      \\ \hline
\defDesktopTwo{}       & NonSitu-Situ    & 2.55     & 2.5    & 1.79     & 36 & 0.29      & 0.61      \\ \hline
\defDesktopTwo{}       & Situ-Situ       & 3.0     & 3.0    & 1.77     & 36 & 0.29      & 0.59      \\ \hline
\defDesktopTwo{}       & Situ-NonSitu    & 2.86     & 3.0    & 1.81     & 36 & 0.30      & 0.61      \\ 
\hline
\hline
\defVR{}       & NonSitu-NonSitu & 3.22     & 3.0    & 1.53     & 36 & 0.26      & 0.52      \\ \hline
\defVR{}      & NonSitu-Situ    & 2.86     & 3.0    & 1.64     & 36 & 0.27      & 0.56      \\ \hline
\defVR{}       & Situ-Situ       & 3.25     & 3.5    & 1.44     & 36 & 0.24      & 0.49      \\ \hline
\defVR{}       & Situ-NonSitu    & 2.53     & 2.0    & 1.75     & 36 & 0.29      & 0.59      \\ \hline
\end{tabu}
\end{table}
\begin{table}[]
\caption{The summary statistics of the accuracy across the \defcongruency{} in \defVR{} study.}
\label{table:acc_congruency_vr}
\footnotesize
\taburulecolor{lightgray}
\begin{tabu}{l|l|l|l|l|l|l|l}

\hline
\textbf{Study} & \textbf{Congruency} & \textbf{mean}     & \textbf{median} & \textbf{sd}       & \textbf{n}  & \textbf{se}        & \textbf{ci}        \\ \hline\hline

\defDesktopOne{}       & Congruent  & 2.72     & 2.5    & 1.87     & 72 & 0.22      & 0.44      \\ \hline
\defDesktopOne{}       & Switch     & 3.06     & 3.0    & 1.74     & 72 & 0.20      & 0.41      \\ \hline


\hline
\defDesktopTwo{}        & Congruent  & 2.88     & 3.0    & 1.84     & 72 & 0.22      & 0.43      \\ \hline
\defDesktopTwo{}         & Switch     & 2.70    & 3.0    & 1.79     & 72 & 0.21    & 0.42      \\ \hline


\hline
VR       & Congruent  & 3.24     & 3.0    & 1.48     & 72 & 0.17      & 0.35      \\ \hline
VR       & Switch     & 2.69     & 3.0    & 1.69     & 72 & 0.20      & 0.40      \\ \hline

\end{tabu}
\end{table}

\subsubsection{All \defModality{} Combined}
The \defModality{} does not affect the memory recall performance (Kruskal-Wallis $\tilde{\chi}^2$ = 0.52745, df = 2, p = 0.7682).
The \defcongruency{} in different \defModality{} does not affect the memory recall performance (Kruskal-Wallis $\tilde{\chi}^2$ = 5.5369, df = 5, p = 0.3539).
The \defgroup{} in different \defModality{} does not affect the memory recall performance (Kruskal-Wallis $\tilde{\chi}^2$ = 8.9419, df = 11, p = 0.6273)

\subsection{Environment Factors}
The environment factor is only available for desktop studies. 

\subsubsection{\defDesktopOne{}}
\begin{figure}[h
b!]
    \centering
    \includegraphics[width=\linewidth]{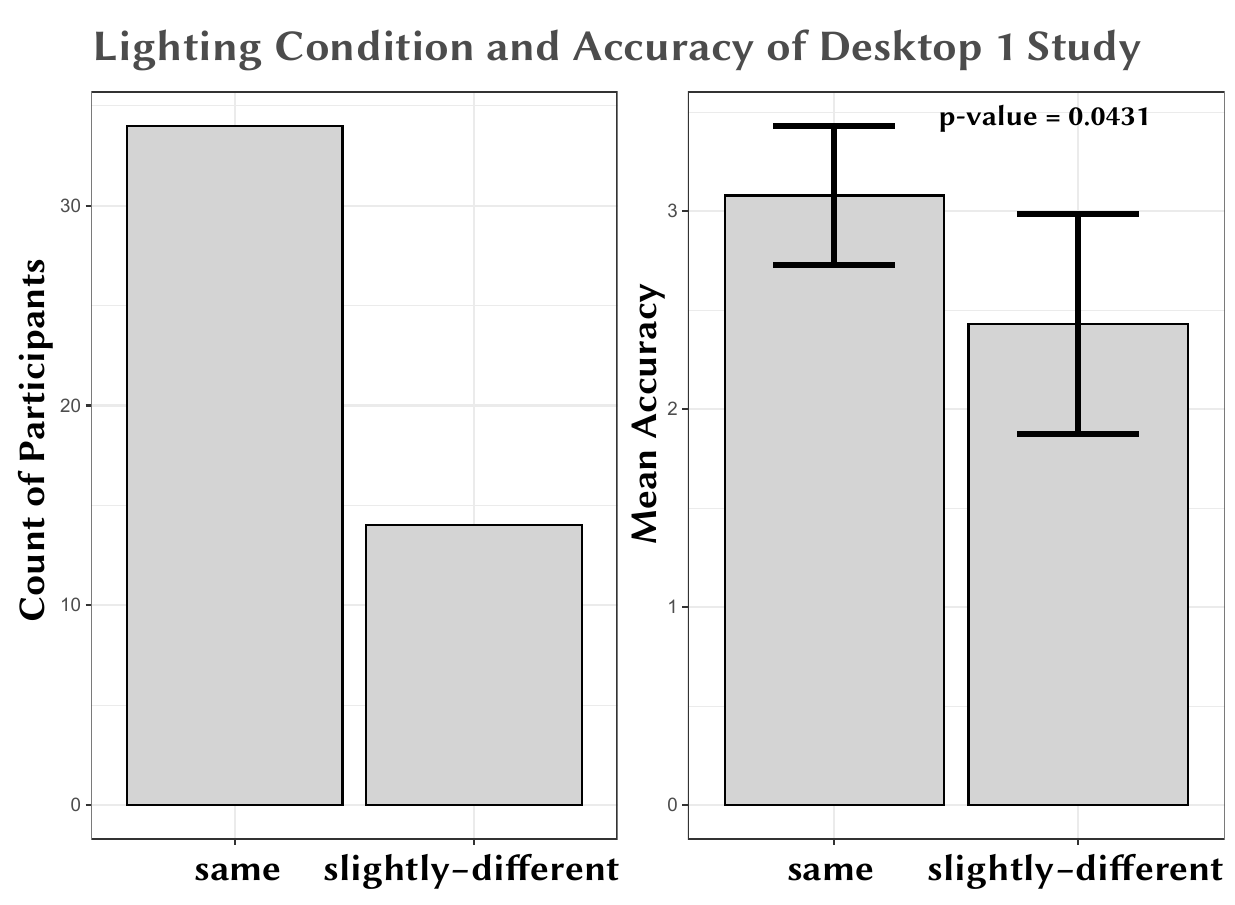}
    \caption{The number of responses per lighting condition (left) and the accuracy (right) in \defDesktopOne. The error bar represents 95\% CI. Same lighting condition average accuracy is 3.08 ($SD$ = 1.79, $CI$ = 0.35), while the slightly different condition accuracy is 2.42 ($SD$ = 1.78, $CI$ = 0.55). 
    Note that an unequal sample size might interfere with the current results.}   \label{fig:plot_lighting_s1}
    \Description{The left figure is a bar chart showing the count of participants in two lighting conditions in study 1. The right figure is a bar chart displaying the recall accuracy of the two lighting conditions.}
\end{figure}

Except the Lighting condition, all factors show that participants were in the same environment conditions during the learning and recall. The lighting condition plot shows that around 14 participants were in slightly different learning and recall environments, while the majority (34 participants) were in the same lighting conditions (Figure \ref{fig:plot_lighting_s1}-left).

We found an indication that the lighting condition between the learning and recall phase might be a significant factor affecting the accuracy (Kruskal-Wallis $\tilde{\chi}^2$ = 4.09, df = 1, p = 0.04314, ANOVA $F$= 4.181, p = 0.042727), as shown in Figure \ref{fig:plot_lighting_s1}-right.

\subsubsection{\defDesktopTwo{}}
All environmental factors including the lighting condition are mainly the same. We could not do a similar analysis as the previous study.

\subsection{Referent Familiarity}
\subsubsection{\defDesktopOne{}}
A little bit more than half of machines (56\% of 72 responses) in the desktop study 1 had never been seen while the remaining 44\% were familiar by the participants.
We did not find supporting evidence that the familiarity of the referent during the learning phase affects the recall accuracy (Kruskal-Wallis $\tilde{\chi}^2$= 0.92555, df = 1, p = 0.336).

\subsubsection{\defDesktopTwo{}}
Almost half (48.6\%) of the referents were never seen by the partiticpants while most of the other half (44\%) were familiar to the participants. A small fraction of machines (6.94\%) were used regularly by the participants.
The familiarity does not affect the recall accuracy (Kruskal-Wallis $\tilde{\chi}^2$= 2.1964, df = 2, p = 0.3335).

\begin{figure}[htb!]
    \centering
    \includegraphics[width=\linewidth]{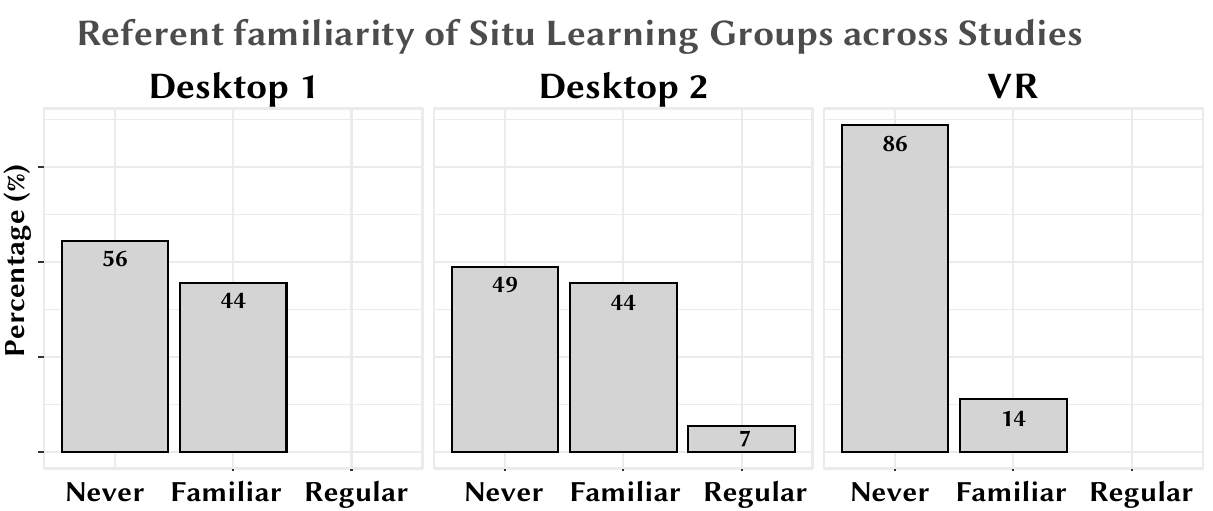}
    \caption{The percentage of referent familiarity levels in all situ groups across \defModality.}   \label{fig:plot_familiarity}
    \Description{The figure includes three bar charts showing the percentage of participants' familiarity with the physical referents in our studies.}
\end{figure}

\subsubsection{\defVR{}} Most participants (86.1\%, 62 out of 72) of the referents were never seen by the partiticpants while the rest (13.8\%) were familiar by the participants.
The familiarity does not affect the recall accuracy (Kruskal-Wallis $\tilde{\chi}^2$= 0.0997, df = 1, p = 0.7522).

\subsection{Referent Order}
The learning order is only available in situ groups. 

\subsubsection{\defDesktopOne{}} 
In the Desktop Study 1, per group test shows that the first learned referent is higher than the last learned in NonSitu-Situ conditon (Friedman $\tilde{\chi}^2$= 9.1351, df = 2, p = 0.01038. ANOVA $F$=3.4742, p = 0.048839).

\subsubsection{\defDesktopTwo{}}
In the Desktop Study 2, per group test shows that the third learned referent is higher than the second learned in Situ-Situ conditon (ANOVA $F$=4.1566, p=0.029423)

\subsubsection{\defVR{}} 
VR We did not find any evidence of the effect of the learning order on the accuracy.

\section{Discussion}
\label{discussion}
\subsection{Summary of results}
The key takeaways from our results are as follows. 

\begin{itemize}[leftmargin=*]
    \item 
The learning processes observed across all the studies yielded primarily successful results, with the majority of participants successfully completing the quizzes in fewer than three attempts. Specifically, in the \defVR{} study, most participants had no prior exposure to the machine, whereas, in the \defDesktopOne{} and \defDesktopTwo{} studies, participants were roughly divided between those unfamiliar with the machine and those already familiar with it. Notably, despite these familiarity-level disparities, our analysis did not reveal any statistically significant impact on accuracy.
 \item
 Within each study, we conducted thorough analyses and found no statistically significant effects on accuracy attributable to the \defgroup{} factor. Similarly, in any of the studies, we did not identify any statistically significant effects associated with the \defcongruency{} factor.
When considering the combined results across all studies, we observed no statistically significant impact of \defModality{} (i.e., the type of experiential situatedness) on accuracy.
Moreover, insights obtained from the \defDesktopOne{} study suggest that the workspace's environment features have an impact. Our findings indicate that participants who experienced consistent lighting conditions during both the learning and recall phases demonstrated significantly higher accuracy than those subjected to slightly differing lighting conditions ($p$ = 0.0431).
\item  In both the \defDesktopTwo{} and \defVR{} studies, a noticeable (but not statistically significant) trend emerges: they appear to align with the CDM effect, wherein recall accuracy is notably higher when learning and recall conditions are congruent compared to non-congruent conditions.
Likewise, the breakdown based on the \defgroup{} factor reveals a parallel pattern. Specifically, recall accuracy in the \defNONsituNONsitu{} condition surpasses that of \defNONsitusitu{}, and accuracy in the \defsitusitu{} condition outperforms \defsituNONsitu{}. 

\item     
In contrast, the \defDesktopOne{} study exhibits a divergent pattern on the \defcongruency{} level. This is primarily attributed to the observation that \defsitusitu{} appears to yield lower recall accuracy compared to \defsituNONsitu{}. 

\item   In both desktop conditions, the presentation sequence appears to have a limited yet notable impact. In the \defDesktopOne{} \defNONsitusitu{} group, we observed that the accuracy of the first referent learned is significantly higher than that of the last one learned ($p$ = 0.048839). Conversely, in the \defDesktopTwo{} \defsitusitu{} group, the third referent learned exhibited a higher level of accuracy compared to the second referent learned ($p$= 0.029423).

\end{itemize}

\subsection{Contextualization of Results}
The context of the information on the desktop screen may not provide sufficient contextual cues for a CDM effect to manifest despite the focused attention on the screen.
In desktop settings, \citet{walti2019reinstating} demonstrated that simple colour and landscape backgrounds could not induce a significant CDM effect for short-term memorization (less than 24 hours) of a list of words. In contrast, our desktop studies investigated long-term memory retention spanning 24 to 48 hours, a timeframe where the potential benefits of CDM had been previously suggested in the literature. Our study's semantic association between the context (physical referent) and the target information (insights) is arguably stronger than that between word lists and landscape pictures in previous studies. However, the mere presence of the physical referent did not significantly impact recall performance when substantial differences were expected. Despite this, the \defDesktopTwo{} study revealed an emerging trend resembling the CDM effect, with congruent conditions yielding slightly higher mean accuracy than the switch conditions.

\citet{mizuho2023virtual} investigated the impact of varying lecturer avatars, serving as contextual elements in lecture videos, on the comprehension of the information presented in the lecture slides. In the control condition, a single avatar was used, while multiple avatars were employed in another condition. The comprehension test, administered within 30 minutes after the lecture, revealed that students exposed to multiple avatar conditions achieved higher comprehension accuracy. This finding resembles the interference reduction effect (see Appendix~\ref{appendixcdm}), where lecture material is divided into multiple segments, each associated with a unique avatar. Although it's acknowledged that the role of context in the study is not definitively established, this work provides deeper insights into CDM studies in desktop settings.
At first glance, our desktop studies may appear similar to \citet{mizuho2023virtual}'s study. Both involve multiple target information items linked to different contexts. However, it's important to note that the lecture material in their study constitutes a single unified piece of target information, meaning its segments are intricately connected to one another. In contrast, the targets in our study can be similar to each other and not necessarily interconnected.

Research has shown that when incidental environmental context cues (unintentional context) are suppressed, the CDM effect is more likely to diminish~\cite{smith2001environmental}. Our underlying assumption regarding desktop environments was that users would concentrate their attention on the information displayed on the screen, effectively suppressing both their workspace context and the environmental proxy context.
The results of our \defDesktopOne{} study indicated that variations in lighting conditions of the user's workspace could impact recall performance. This phenomenon was observed when participants in our study reported that their workspaces mostly remained the same during both the learning and recall phases, with only slight differences in lighting conditions. Our initial speculation is that the altered lighting conditions within the workspace may have been perceived as a change in context by the participants before they entered the recall phase. Consequently, this perceived change minimized the influence of suppressed environmental cues during the actual recall test, leading to differences in recall performance that mirror CDM effects. 

It's challenging to clarify why the \defsitusitu{} group appears to exhibit lower performance compared to the \defsituNONsitu{} group in the \defDesktopOne{} study. Our examination of the lighting conditions in the \defsitusitu{} group revealed no discernible patterns (see the analysis in the Supplementary Material).
We can ascertain that only the \defDesktopOne{} study involves loosely associated target information, as depicted in Figure \ref{fig:conversion}. The potential impact of such a variation in data representation, if indeed the root cause, remains unclear.

Previous CDM studies conducted in VR have explored diverse virtual environments, including Mars vs. Underwater~\cite{shin2021context}, Mountain vs. Underwater~\cite{de2018applicability}, Fantasy Worlds~\cite{essoe2022enhancing}, or Indoor vs. Outdoor virtual environments~\cite{chocholavckova2023context}. To some extent, the results of these studies show the CDM effect for long-term recall but not short-term recall. They operate on the premise that the entire virtual scene serves as context, illustrating that manipulating the learning and recall environments can trigger CDM effects. 
In our VR study, we maintained consistent environments around the user and the physical referent (in situ-learning conditions) to investigate whether the presence of physical object proxies closely related to target information could induce CDM effects. We did not identify statistically significant evidence of CDM effects. However, we did observe that congruent conditions (\defsitusitu{} and \defNONsituNONsitu{}) exhibited higher mean accuracy compared to switch conditions (\defsituNONsitu{} and \defNONsitusitu{}). This trend appears more pronounced in VR compared to desktop settings. This suggests that while the presence of physical referent proxies in immersive environments may offer a contextual memory cue, the effect size may be smaller than anticipated based on our study's statistical parameters. Hence, revealing a smaller effect size requires significantly more samples. 

\subsection{Reflection on Situated Visualization}

\textbf{The inclusion of a referent during recall may result in minor disruptions in non-situated learning}. Non-situated learning accuracy declines when a physical referent is present during recall. In the \defDesktopTwo{} study, the average difference between \defNONsituNONsitu{} and \defNONsitusitu{} is 0.23 points, equivalent to an 8.27\% decrease from the congruent context. Meanwhile, in the \defVR{} study, the difference is 0.36, representing an 11.2\% decrease from the congruent context.
In the desktop study, most participants were in the same workspace condition, and the environment surrounding the physical referent (i.e., the white background) remained consistent. Similarly, in the \defVR{} study, the perceived environment by the participant (an empty scene) remained unchanged. This suggests that despite the reinstatement of the same learning context, the presence of the referent can still lead to a slight disruption.

\textbf{The absence of the referent can have a minor detrimental impact on recall in situated learning}. Situated learning accuracy declines when the physical referent is missing during recall. In \defDesktopTwo{}, the average difference between \defsitusitu{}{} and \defsituNONsitu{} is 0.14 points, representing a 4.7\% decrease from the congruent context. In the \defVR{} study, this difference increases to 0.72, marking a 22.1\% decrease from the congruent context.
During the learning stage, where participants contextualise physical referents, they may naturally anticipate encountering these contexts during recall. The absence of the referents can then lead to a contextual mismatch, slightly impacting memory recall.

\textbf{In VR environments, the significance of the referent's presence may be less pronounced than the congruency between the learning and recall environments}. The difference in accuracy between \defNONsituNONsitu{} and \defsitusitu{} (congruent context) is a modest 0.03, which pales in comparison to the differences observed when the learning context undergoes a shift (ranging from 0.14 to 0.72). This suggests that prioritizing the consistency of the learning context might take precedence over ensuring the presence of a physical referent.
For example, in scenarios where users acquire knowledge through abstract representations like those found in Imaxes~\cite{cordeil2017imaxes}, introducing a representation of the physical referent may even harm memorability.

\textbf{What can we learn from the impacts on recall accuracy?}
In our study, recall accuracy measures the participant's ability to recall attribute-pattern pairs correctly. Across our three studies, average accuracy ranged from 2.47 to 3.25. A score of 3 signifies an accurate recall of the attribute and its associated pattern (e.g., ``I only remember the breakdown is increasing''). A score of 2 indicates correct attribute recall but an inability to describe its characteristics (e.g., ``the information related to breakdown and performance, but I am uncertain about the patterns'').
Beyond the semantics of these pairs, accuracy also reflects the likelihood of a correct answer from a finite set of options. An accuracy score of 3 corresponds to a 41.7\% chance of providing a correct response (1/4 + 1/6), while an accuracy score of 2 reflects a 50\% chance of being correct (1/4 + 1/4).
The significance of the observed variation in accuracy between the minimum and maximum scores is open to interpretation and may vary in different use cases. The absence of a statistically significant effect suggests that, if a CDM effect exists, it may be smaller than anticipated in our study. The practical implications of such a difference in situated visualization remain a subject for future research.

\textbf{Immersion does not appear to enhance recall ability, but it may intensify sensitivity to context switching effects}. 
Immersive VR environments offer the advantage of audio-visual isolation, potentially reducing distractions during data analysis. Additionally, VR proxsituated visualization can create a strong sense of presence in immediate surroundings. However, this heightened isolation and presence may amplify the influence of contextual cues, potentially impacting memory recall performance.
In our studies, the presence or absence of the proxy in VR seems to induce more significant context switching compared to desktop studies, where the presence or absence of the physical referent represents a subtler contextual change. In \defDesktopTwo{}, the difference between average accuracy in congruent (2.88) and switch (2.70) conditions is 0.18 points, equivalent to 6.66\% of the switch mean accuracy. In the \defVR{} study, this difference between congruent (3.24) and switch (2.69) conditions is 0.55, accounting for 20.4\% of the switch mean accuracy.
This explains the observed trend in VR, where switch groups exhibit lower accuracy than congruent groups. These findings prompt questions about the advantages and disadvantages of CDM in situated visualization, particularly in understanding when contextual dependency aids or hinders performance. Further research is needed to shed light on these aspects.

\textbf{If preserving the surrounding context is a priority, careful consideration may be required when using a fully immersive referent proxy}. In the near future, as immersive 'spatial' computing becomes more prevalent, integrating immersive proxy representations as ad-hoc contexts in non-situated data visualization is conceivable. However, when the analytical process doesn't heavily rely on physical referent representation, or it's only occasionally necessary, the proxy presentation should be subtle, especially if memorability is a concern.
For instance, envision a scenario where a user has been exploring a dataset on a desktop, and they receive a 3D model update of physical referents from a colleague. Instead of transitioning into a fully immersive VR experience, these referents could be subtly integrated into the environment through small-scale virtual proxies in augmented reality, positioned on the desk near the associated charts.

\section{Limitations and Future Work}
\label{limitations}
We recognize several limitations within our study. Firstly, expanding the participant pool in a virtual reality setting would benefit future research, as it could bolster the emerging patterns we observed. Secondly, while the crowd-sourcing service employed for our desktop studies is considered one of the best available options, we acknowledge that the studies were not conducted within a tightly controlled environment. Thirdly, memory studies often employ various methods to measure memory recall accuracy, but our study is confined to a discrete, ordinal measure.

Future studies could explore alternative methods, such as free recall of insights. Additionally, internal factors hold relevance; for instance, research has shown that stress can diminish the CDM effect~\cite{schwabe2009stress}. Investigating how these internal factors relate to situated visualization represents a valuable avenue for future research.

In our study, the learning stage followed a discrete and consecutive approach, aligning with navigation methods like click-to-navigate or teleportation in VR. However, certain scenarios involve continuous navigation techniques, such as flyover or physical locomotion, during data exploration. In these cases, spatial awareness and spatial memory may contribute additional context. Exploring whether these variations in data exploration techniques during the learning stage yield different outcomes is a question worthy of exploration. Ultimately, a vast unknown space for future research is delineated within our design framework. Examples of such research areas encompass the investigation of a single referent, multiple homogeneous referents, different types of data representations, and determining the threshold for when the number of referents or complexity of data representation becomes excessive. Moreover, cross-experiences hold significant relevance for the future. For instance, consider scenarios, where learning occurs in an immediate situated visualization but recall, takes place in a proxsituated on a desktop or in VR.

\section{Conclusion}
In this paper, we leverage the concept of context-dependent memory from cognitive psychology to assess situated visualization. We acknowledge similarities and differences between previous studies on context-dependent memory and scenarios involving situated visualization. We propose a list of crucial factors, outlining a design space with twelve distinct contextual scenarios. We explored proxsituated visualization scenarios in both desktop and VR environments, employing diverse contextual referents coupled with textual insights as memory targets. Our research has revealed that relying solely on the proxy referent may not significantly induce context-dependent memory (CDM) effects, whereas the workspace (for desktop settings) and proxy environment (for VR settings) could have more pronounced contextual cues. 
This represents an initial step towards gaining a deeper understanding of cognitive processes in situated visualization.


\bibliographystyle{ACM-Reference-Format}
\bibliography{main}


\begin{thebibliography}{62}


\ifx \showCODEN    \undefined \def \showCODEN     #1{\unskip}     \fi
\ifx \showDOI      \undefined \def \showDOI       #1{#1}\fi
\ifx \showISBNx    \undefined \def \showISBNx     #1{\unskip}     \fi
\ifx \showISBNxiii \undefined \def \showISBNxiii  #1{\unskip}     \fi
\ifx \showISSN     \undefined \def \showISSN      #1{\unskip}     \fi
\ifx \showLCCN     \undefined \def \showLCCN      #1{\unskip}     \fi
\ifx \shownote     \undefined \def \shownote      #1{#1}          \fi
\ifx \showarticletitle \undefined \def \showarticletitle #1{#1}   \fi
\ifx \showURL      \undefined \def \showURL       {\relax}        \fi
\providecommand\bibfield[2]{#2}
\providecommand\bibinfo[2]{#2}
\providecommand\natexlab[1]{#1}
\providecommand\showeprint[2][]{arXiv:#2}

\bibitem[Alonso et~al\mbox{.}(2018)]%
        {alonso2018cityscope}
\bibfield{author}{\bibinfo{person}{Luis Alonso}, \bibinfo{person}{Yan~Ryan Zhang}, \bibinfo{person}{Arnaud Grignard}, \bibinfo{person}{Ariel Noyman}, \bibinfo{person}{Yasushi Sakai}, \bibinfo{person}{Markus ElKatsha}, \bibinfo{person}{Ronan Doorley}, {and} \bibinfo{person}{Kent Larson}.} \bibinfo{year}{2018}\natexlab{}.
\newblock \showarticletitle{Cityscope: a data-driven interactive simulation tool for urban design. Use case volpe}. In \bibinfo{booktitle}{\emph{Unifying Themes in Complex Systems IX: Proceedings of the Ninth International Conference on Complex Systems 9}}. Springer, \bibinfo{pages}{253--261}.
\newblock


\bibitem[Assor et~al\mbox{.}(2023)]%
        {assor2023handling}
\bibfield{author}{\bibinfo{person}{Ambre Assor}, \bibinfo{person}{Arnaud Prouzeau}, \bibinfo{person}{Martin Hachet}, {and} \bibinfo{person}{Pierre Dragicevic}.} \bibinfo{year}{2023}\natexlab{}.
\newblock \showarticletitle{Handling Non-Visible Referents in Situated Visualizations}.
\newblock  (\bibinfo{year}{2023}).
\newblock


\bibitem[Bateman et~al\mbox{.}(2010)]%
        {bateman2010useful}
\bibfield{author}{\bibinfo{person}{Scott Bateman}, \bibinfo{person}{Regan~L Mandryk}, \bibinfo{person}{Carl Gutwin}, \bibinfo{person}{Aaron Genest}, \bibinfo{person}{David McDine}, {and} \bibinfo{person}{Christopher Brooks}.} \bibinfo{year}{2010}\natexlab{}.
\newblock \showarticletitle{Useful junk? The effects of visual embellishment on comprehension and memorability of charts}. In \bibinfo{booktitle}{\emph{Proceedings of the SIGCHI conference on human factors in computing systems}}. \bibinfo{pages}{2573--2582}.
\newblock


\bibitem[Bonnail et~al\mbox{.}(2023)]%
        {bonnail2023memory}
\bibfield{author}{\bibinfo{person}{Elise Bonnail}, \bibinfo{person}{Wen-Jie Tseng}, \bibinfo{person}{Mark McGill}, \bibinfo{person}{Eric Lecolinet}, \bibinfo{person}{Samuel Huron}, {and} \bibinfo{person}{Jan Gugenheimer}.} \bibinfo{year}{2023}\natexlab{}.
\newblock \showarticletitle{Memory Manipulations in Extended Reality}. In \bibinfo{booktitle}{\emph{Proceedings of the 2023 CHI Conference on Human Factors in Computing Systems}}. \bibinfo{pages}{1--20}.
\newblock


\bibitem[Borkin et~al\mbox{.}(2013)]%
        {borkin2013makes}
\bibfield{author}{\bibinfo{person}{Michelle~A Borkin}, \bibinfo{person}{Azalea~A Vo}, \bibinfo{person}{Zoya Bylinskii}, \bibinfo{person}{Phillip Isola}, \bibinfo{person}{Shashank Sunkavalli}, \bibinfo{person}{Aude Oliva}, {and} \bibinfo{person}{Hanspeter Pfister}.} \bibinfo{year}{2013}\natexlab{}.
\newblock \showarticletitle{What makes a visualization memorable?}
\newblock \bibinfo{journal}{\emph{IEEE transactions on visualization and computer graphics}} \bibinfo{volume}{19}, \bibinfo{number}{12} (\bibinfo{year}{2013}), \bibinfo{pages}{2306--2315}.
\newblock


\bibitem[Bressa et~al\mbox{.}(2021)]%
        {bressa2021s}
\bibfield{author}{\bibinfo{person}{Nathalie Bressa}, \bibinfo{person}{Henrik Korsgaard}, \bibinfo{person}{Aur{\'e}lien Tabard}, \bibinfo{person}{Steven Houben}, {and} \bibinfo{person}{Jo Vermeulen}.} \bibinfo{year}{2021}\natexlab{}.
\newblock \showarticletitle{What's the situation with situated visualization? A survey and perspectives on situatedness}.
\newblock \bibinfo{journal}{\emph{IEEE Transactions on Visualization and Computer Graphics}} \bibinfo{volume}{28}, \bibinfo{number}{1} (\bibinfo{year}{2021}), \bibinfo{pages}{107--117}.
\newblock


\bibitem[Cain and Still(2019)]%
        {cain2019graphical}
\bibfield{author}{\bibinfo{person}{Ashley~A Cain} {and} \bibinfo{person}{Jeremiah~D Still}.} \bibinfo{year}{2019}\natexlab{}.
\newblock \showarticletitle{Graphical Authentication Passcode Memorability: Context, Length, and Number}. In \bibinfo{booktitle}{\emph{Proceedings of the Human Factors and Ergonomics Society Annual Meeting}}, Vol.~\bibinfo{volume}{63}. SAGE Publications Sage CA: Los Angeles, CA, \bibinfo{pages}{447--451}.
\newblock


\bibitem[Chochol{\'a}{\v{c}}kov{\'a} et~al\mbox{.}(2023)]%
        {chocholavckova2023context}
\bibfield{author}{\bibinfo{person}{M{\'a}ria Chochol{\'a}{\v{c}}kov{\'a}}, \bibinfo{person}{Vojt{\v{e}}ch Ju{\v{r}}{\'\i}k}, \bibinfo{person}{Alexandra Ru{\v{z}}i{\v{c}}kov{\'a}}, \bibinfo{person}{Lenka Jurkovi{\v{c}}ov{\'a}}, \bibinfo{person}{Pavel Ugwitz}, {and} \bibinfo{person}{Martin Jel{\'\i}nek}.} \bibinfo{year}{2023}\natexlab{}.
\newblock \showarticletitle{Context-dependent memory recall in HMD-based immersive virtual environments}.
\newblock \bibinfo{journal}{\emph{Plos one}} \bibinfo{volume}{18}, \bibinfo{number}{8} (\bibinfo{year}{2023}), \bibinfo{pages}{e0289079}.
\newblock


\bibitem[Collins et~al\mbox{.}(2020)]%
        {collins2020augmented}
\bibfield{author}{\bibinfo{person}{Toby Collins}, \bibinfo{person}{Daniel Pizarro}, \bibinfo{person}{Simone Gasparini}, \bibinfo{person}{Nicolas Bourdel}, \bibinfo{person}{Pauline Chauvet}, \bibinfo{person}{Michel Canis}, \bibinfo{person}{Lilian Calvet}, {and} \bibinfo{person}{Adrien Bartoli}.} \bibinfo{year}{2020}\natexlab{}.
\newblock \showarticletitle{Augmented reality guided laparoscopic surgery of the uterus}.
\newblock \bibinfo{journal}{\emph{IEEE Transactions on Medical Imaging}} \bibinfo{volume}{40}, \bibinfo{number}{1} (\bibinfo{year}{2020}), \bibinfo{pages}{371--380}.
\newblock


\bibitem[Cordeil et~al\mbox{.}(2017)]%
        {cordeil2017imaxes}
\bibfield{author}{\bibinfo{person}{Maxime Cordeil}, \bibinfo{person}{Andrew Cunningham}, \bibinfo{person}{Tim Dwyer}, \bibinfo{person}{Bruce~H Thomas}, {and} \bibinfo{person}{Kim Marriott}.} \bibinfo{year}{2017}\natexlab{}.
\newblock \showarticletitle{ImAxes: Immersive axes as embodied affordances for interactive multivariate data visualisation}. In \bibinfo{booktitle}{\emph{Proceedings of the 30th annual ACM symposium on user interface software and technology}}. \bibinfo{pages}{71--83}.
\newblock


\bibitem[Cunningham et~al\mbox{.}(2021)]%
        {cunningham2021towards}
\bibfield{author}{\bibinfo{person}{Andrew Cunningham}, \bibinfo{person}{Jonathon~Derek Hart}, \bibinfo{person}{Ulrich Engelke}, \bibinfo{person}{Matt Adcock}, {and} \bibinfo{person}{Bruce~H Thomas}.} \bibinfo{year}{2021}\natexlab{}.
\newblock \showarticletitle{Towards Embodied Interaction for Geospatial Energy Sector Analytics in Immersive Environments}. In \bibinfo{booktitle}{\emph{Proceedings of the Twelfth ACM International Conference on Future Energy Systems}}. \bibinfo{pages}{396--400}.
\newblock


\bibitem[de~Back et~al\mbox{.}(2018)]%
        {de2018applicability}
\bibfield{author}{\bibinfo{person}{Tycho~T de Back}, \bibinfo{person}{Angelica~M Tinga}, \bibinfo{person}{Rens van Hoef}, \bibinfo{person}{Erwin~M Peters}, {and} \bibinfo{person}{Max~M Louwerse}.} \bibinfo{year}{2018}\natexlab{}.
\newblock \showarticletitle{The applicability and benefits of virtual reality for the cognitive sciences: The case of context-dependent memory}. In \bibinfo{booktitle}{\emph{Proceedings of the 40th Annual Conference of the Cognitive Science Society}}. \bibinfo{pages}{293--298}.
\newblock


\bibitem[Dudai et~al\mbox{.}(2007)]%
        {dudai2007memory}
\bibfield{author}{\bibinfo{person}{Yadin Dudai}, \bibinfo{person}{HL Roediger~Ill}, \bibinfo{person}{Endel Tulving}, \bibinfo{person}{HL Roediger~III}, \bibinfo{person}{Y Dudai}, {and} \bibinfo{person}{SM Fitzpatrick}.} \bibinfo{year}{2007}\natexlab{}.
\newblock \showarticletitle{Memory concepts}.
\newblock \bibinfo{journal}{\emph{Science of memory: Concepts}} (\bibinfo{year}{2007}), \bibinfo{pages}{1--9}.
\newblock


\bibitem[Ellenberg et~al\mbox{.}(2023)]%
        {ellenberg2023spatiality}
\bibfield{author}{\bibinfo{person}{Mats~Ole Ellenberg}, \bibinfo{person}{Marc Satkowski}, \bibinfo{person}{Weizhou Luo}, {and} \bibinfo{person}{Raimund Dachselt}.} \bibinfo{year}{2023}\natexlab{}.
\newblock \showarticletitle{Spatiality and Semantics-Towards Understanding Content Placement in Mixed Reality}. In \bibinfo{booktitle}{\emph{Extended Abstracts of the 2023 CHI Conference on Human Factors in Computing Systems}}. \bibinfo{pages}{1--8}.
\newblock


\bibitem[ElSayed et~al\mbox{.}(2015)]%
        {elsayed2015situated}
\bibfield{author}{\bibinfo{person}{Neven ElSayed}, \bibinfo{person}{Bruce Thomas}, \bibinfo{person}{Kim Marriott}, \bibinfo{person}{Julia Piantadosi}, {and} \bibinfo{person}{Ross Smith}.} \bibinfo{year}{2015}\natexlab{}.
\newblock \showarticletitle{Situated analytics}. In \bibinfo{booktitle}{\emph{2015 Big Data Visual Analytics (BDVA)}}. IEEE, \bibinfo{pages}{1--8}.
\newblock


\bibitem[Elsayed et~al\mbox{.}(2015)]%
        {elsayed2015using}
\bibfield{author}{\bibinfo{person}{Neven~AM Elsayed}, \bibinfo{person}{Bruce~H Thomas}, \bibinfo{person}{Ross~T Smith}, \bibinfo{person}{Kim Marriott}, {and} \bibinfo{person}{Julia Piantadosi}.} \bibinfo{year}{2015}\natexlab{}.
\newblock \showarticletitle{Using augmented reality to support situated analytics}. In \bibinfo{booktitle}{\emph{2015 IEEE Virtual Reality (VR)}}. IEEE, \bibinfo{pages}{175--176}.
\newblock


\bibitem[Ens et~al\mbox{.}(2021)]%
        {ens2021grand}
\bibfield{author}{\bibinfo{person}{Barrett Ens}, \bibinfo{person}{Benjamin Bach}, \bibinfo{person}{Maxime Cordeil}, \bibinfo{person}{Ulrich Engelke}, \bibinfo{person}{Marcos Serrano}, \bibinfo{person}{Wesley Willett}, \bibinfo{person}{Arnaud Prouzeau}, \bibinfo{person}{Christoph Anthes}, \bibinfo{person}{Wolfgang B{\"u}schel}, \bibinfo{person}{Cody Dunne}, {et~al\mbox{.}}} \bibinfo{year}{2021}\natexlab{}.
\newblock \showarticletitle{Grand challenges in immersive analytics}. In \bibinfo{booktitle}{\emph{Proceedings of the 2021 CHI Conference on Human Factors in Computing Systems}}. \bibinfo{pages}{1--17}.
\newblock


\bibitem[Ens et~al\mbox{.}(2020)]%
        {ens2020uplift}
\bibfield{author}{\bibinfo{person}{Barrett Ens}, \bibinfo{person}{Sarah Goodwin}, \bibinfo{person}{Arnaud Prouzeau}, \bibinfo{person}{Fraser Anderson}, \bibinfo{person}{Florence~Y Wang}, \bibinfo{person}{Samuel Gratzl}, \bibinfo{person}{Zac Lucarelli}, \bibinfo{person}{Brendan Moyle}, \bibinfo{person}{Jim Smiley}, {and} \bibinfo{person}{Tim Dwyer}.} \bibinfo{year}{2020}\natexlab{}.
\newblock \showarticletitle{Uplift: A tangible and immersive tabletop system for casual collaborative visual analytics}.
\newblock \bibinfo{journal}{\emph{IEEE Transactions on Visualization and Computer Graphics}} \bibinfo{volume}{27}, \bibinfo{number}{2} (\bibinfo{year}{2020}), \bibinfo{pages}{1193--1203}.
\newblock


\bibitem[Essoe et~al\mbox{.}(2022)]%
        {essoe2022enhancing}
\bibfield{author}{\bibinfo{person}{Joey Ka-Yee Essoe}, \bibinfo{person}{Nicco Reggente}, \bibinfo{person}{Ai~Aileen Ohno}, \bibinfo{person}{Younji~Hera Baek}, \bibinfo{person}{John Dell’Italia}, {and} \bibinfo{person}{Jesse Rissman}.} \bibinfo{year}{2022}\natexlab{}.
\newblock \showarticletitle{Enhancing learning and retention with distinctive virtual reality environments and mental context reinstatement}.
\newblock \bibinfo{journal}{\emph{npj Science of Learning}} \bibinfo{volume}{7}, \bibinfo{number}{1} (\bibinfo{year}{2022}), \bibinfo{pages}{31}.
\newblock


\bibitem[Fleck et~al\mbox{.}(2022)]%
        {fleck2022ragrug}
\bibfield{author}{\bibinfo{person}{Philipp Fleck}, \bibinfo{person}{Aimee~Sousa Calepso}, \bibinfo{person}{Sebastian Hubenschmid}, \bibinfo{person}{Michael Sedlmair}, {and} \bibinfo{person}{Dieter Schmalstieg}.} \bibinfo{year}{2022}\natexlab{}.
\newblock \showarticletitle{Ragrug: A toolkit for situated analytics}.
\newblock \bibinfo{journal}{\emph{IEEE Transactions on Visualization and Computer Graphics}} (\bibinfo{year}{2022}).
\newblock


\bibitem[Godden and Baddeley(1975)]%
        {godden1975context}
\bibfield{author}{\bibinfo{person}{Duncan~R Godden} {and} \bibinfo{person}{Alan~D Baddeley}.} \bibinfo{year}{1975}\natexlab{}.
\newblock \showarticletitle{Context-dependent memory in two natural environments: On land and underwater}.
\newblock \bibinfo{journal}{\emph{British Journal of psychology}} \bibinfo{volume}{66}, \bibinfo{number}{3} (\bibinfo{year}{1975}), \bibinfo{pages}{325--331}.
\newblock


\bibitem[Guarese et~al\mbox{.}(2020)]%
        {guarese2020augmented}
\bibfield{author}{\bibinfo{person}{Renan Guarese}, \bibinfo{person}{Jo{\~a}o Becker}, \bibinfo{person}{Henrique Fensterseifer}, \bibinfo{person}{Marcelo Walter}, \bibinfo{person}{Carla Freitas}, \bibinfo{person}{Luciana Nedel}, {and} \bibinfo{person}{Anderson Maciel}.} \bibinfo{year}{2020}\natexlab{}.
\newblock \showarticletitle{Augmented situated visualization for spatial and context-aware decision-making}. In \bibinfo{booktitle}{\emph{Proceedings of the International Conference on Advanced Visual Interfaces}}. \bibinfo{pages}{1--5}.
\newblock


\bibitem[Harman et~al\mbox{.}(2019)]%
        {harman2019role}
\bibfield{author}{\bibinfo{person}{Joel Harman}, \bibinfo{person}{Ross Brown}, \bibinfo{person}{Daniel Johnson}, {and} \bibinfo{person}{Selen Turkay}.} \bibinfo{year}{2019}\natexlab{}.
\newblock \showarticletitle{The role of visual detail during situated memory recall within a virtual reality environment}. In \bibinfo{booktitle}{\emph{Proceedings of the 31st Australian Conference on Human-computer-Interaction}}. \bibinfo{pages}{138--148}.
\newblock


\bibitem[Krokos et~al\mbox{.}(2019)]%
        {krokos2019virtual}
\bibfield{author}{\bibinfo{person}{Eric Krokos}, \bibinfo{person}{Catherine Plaisant}, {and} \bibinfo{person}{Amitabh Varshney}.} \bibinfo{year}{2019}\natexlab{}.
\newblock \showarticletitle{Virtual memory palaces: immersion aids recall}.
\newblock \bibinfo{journal}{\emph{Virtual reality}}  \bibinfo{volume}{23} (\bibinfo{year}{2019}), \bibinfo{pages}{1--15}.
\newblock


\bibitem[Lamers and Lanen(2021)]%
        {lamers2021changing}
\bibfield{author}{\bibinfo{person}{Maarten~H Lamers} {and} \bibinfo{person}{Maik Lanen}.} \bibinfo{year}{2021}\natexlab{}.
\newblock \showarticletitle{Changing between virtual reality and real-world adversely affects memory recall accuracy}.
\newblock \bibinfo{journal}{\emph{Frontiers in Virtual Reality}}  \bibinfo{volume}{2} (\bibinfo{year}{2021}), \bibinfo{pages}{602087}.
\newblock


\bibitem[Lee et~al\mbox{.}(2023)]%
        {lee2023design}
\bibfield{author}{\bibinfo{person}{Benjamin Lee}, \bibinfo{person}{Michael Sedlmair}, {and} \bibinfo{person}{Dieter Schmalstieg}.} \bibinfo{year}{2023}\natexlab{}.
\newblock \showarticletitle{Design Patterns for Situated Visualization in Augmented Reality}.
\newblock \bibinfo{journal}{\emph{arXiv preprint arXiv:2307.09157}} (\bibinfo{year}{2023}).
\newblock


\bibitem[Lee et~al\mbox{.}(2013)]%
        {lee2013effects}
\bibfield{author}{\bibinfo{person}{Cha Lee}, \bibinfo{person}{Gustavo~A Rincon}, \bibinfo{person}{Greg Meyer}, \bibinfo{person}{Tobias H{\"o}llerer}, {and} \bibinfo{person}{Doug~A Bowman}.} \bibinfo{year}{2013}\natexlab{}.
\newblock \showarticletitle{The effects of visual realism on search tasks in mixed reality simulation}.
\newblock \bibinfo{journal}{\emph{IEEE transactions on visualization and computer graphics}} \bibinfo{volume}{19}, \bibinfo{number}{4} (\bibinfo{year}{2013}), \bibinfo{pages}{547--556}.
\newblock


\bibitem[Lin et~al\mbox{.}(2022)]%
        {lin2022quest}
\bibfield{author}{\bibinfo{person}{Tica Lin}, \bibinfo{person}{Zhutian Chen}, \bibinfo{person}{Yalong Yang}, \bibinfo{person}{Daniele Chiappalupi}, \bibinfo{person}{Johanna Beyer}, {and} \bibinfo{person}{Hanspeter Pfister}.} \bibinfo{year}{2022}\natexlab{}.
\newblock \showarticletitle{The Quest for: Embedded Visualization for Augmenting Basketball Game Viewing Experiences}.
\newblock \bibinfo{journal}{\emph{IEEE transactions on visualization and computer graphics}} \bibinfo{volume}{29}, \bibinfo{number}{1} (\bibinfo{year}{2022}), \bibinfo{pages}{962--971}.
\newblock


\bibitem[Liu et~al\mbox{.}(2022)]%
        {liu2022effects}
\bibfield{author}{\bibinfo{person}{Jiazhou Liu}, \bibinfo{person}{Arnaud Prouzeau}, \bibinfo{person}{Barrett Ens}, {and} \bibinfo{person}{Tim Dwyer}.} \bibinfo{year}{2022}\natexlab{}.
\newblock \showarticletitle{Effects of Display Layout on Spatial Memory for Immersive Environments}.
\newblock \bibinfo{journal}{\emph{Proceedings of the ACM on Human-Computer Interaction}} \bibinfo{volume}{6}, \bibinfo{number}{ISS} (\bibinfo{year}{2022}), \bibinfo{pages}{468--488}.
\newblock


\bibitem[Mizuho et~al\mbox{.}(2023a)]%
        {mizuho2023virtual}
\bibfield{author}{\bibinfo{person}{Takato Mizuho}, \bibinfo{person}{Tomohiro Amemiya}, \bibinfo{person}{Takuji Narumi}, {and} \bibinfo{person}{Hideaki Kuzuoka}.} \bibinfo{year}{2023}\natexlab{a}.
\newblock \showarticletitle{Virtual Omnibus Lecture: Investigating the Effects of Varying Lecturer Avatars as Environmental Context on Audience Memory}. In \bibinfo{booktitle}{\emph{Proceedings of the Augmented Humans International Conference 2023}}. \bibinfo{pages}{55--65}.
\newblock


\bibitem[Mizuho et~al\mbox{.}(2023b)]%
        {mizuho2023effects}
\bibfield{author}{\bibinfo{person}{Takato Mizuho}, \bibinfo{person}{Takuji Narumi}, {and} \bibinfo{person}{Hideaki Kuzuoka}.} \bibinfo{year}{2023}\natexlab{b}.
\newblock \showarticletitle{Effects of the Visual Fidelity of Virtual Environments on Presence, Context-dependent Forgetting, and Source-monitoring Error}.
\newblock \bibinfo{journal}{\emph{IEEE Transactions on Visualization and Computer Graphics}} \bibinfo{volume}{29}, \bibinfo{number}{5} (\bibinfo{year}{2023}), \bibinfo{pages}{2607--2614}.
\newblock


\bibitem[Moere and Hill(2012)]%
        {moere2012designing}
\bibfield{author}{\bibinfo{person}{Andrew~Vande Moere} {and} \bibinfo{person}{Dan Hill}.} \bibinfo{year}{2012}\natexlab{}.
\newblock \showarticletitle{Designing for the situated and public visualization of urban data}.
\newblock \bibinfo{journal}{\emph{Journal of Urban Technology}} \bibinfo{volume}{19}, \bibinfo{number}{2} (\bibinfo{year}{2012}), \bibinfo{pages}{25--46}.
\newblock


\bibitem[Parker et~al\mbox{.}(2020)]%
        {parker2020exploring}
\bibfield{author}{\bibinfo{person}{Jason~A Parker}, \bibinfo{person}{Alexandra~D Kaplan}, \bibinfo{person}{William~G Volante}, \bibinfo{person}{Julian Abich~IV}, {and} \bibinfo{person}{Valerie~K Sims}.} \bibinfo{year}{2020}\natexlab{}.
\newblock \showarticletitle{Exploring the encoding specificity principle and context-dependent recognition in virtual reality}. In \bibinfo{booktitle}{\emph{Proceedings of the Human Factors and Ergonomics Society Annual Meeting}}, Vol.~\bibinfo{volume}{64}. SAGE Publications Sage CA: Los Angeles, CA, \bibinfo{pages}{1481--1485}.
\newblock


\bibitem[Plancher et~al\mbox{.}(2010)]%
        {plancher2010age}
\bibfield{author}{\bibinfo{person}{Ga{\"e}n Plancher}, \bibinfo{person}{Valerie Gyselinck}, \bibinfo{person}{Serge Nicolas}, {and} \bibinfo{person}{Pascale Piolino}.} \bibinfo{year}{2010}\natexlab{}.
\newblock \showarticletitle{Age effect on components of episodic memory and feature binding: A virtual reality study.}
\newblock \bibinfo{journal}{\emph{Neuropsychology}} \bibinfo{volume}{24}, \bibinfo{number}{3} (\bibinfo{year}{2010}), \bibinfo{pages}{379}.
\newblock


\bibitem[Pooryousef et~al\mbox{.}(2023)]%
        {pooryousef2023working}
\bibfield{author}{\bibinfo{person}{Vahid Pooryousef}, \bibinfo{person}{Maxime Cordeil}, \bibinfo{person}{Lonni Besan{\c{c}}on}, \bibinfo{person}{Christophe Hurter}, \bibinfo{person}{Tim Dwyer}, {and} \bibinfo{person}{Richard Bassed}.} \bibinfo{year}{2023}\natexlab{}.
\newblock \showarticletitle{Working with Forensic Practitioners to Understand the Opportunities and Challenges for Mixed-Reality Digital Autopsy}. In \bibinfo{booktitle}{\emph{Proceedings of the 2023 CHI Conference on Human Factors in Computing Systems}}. \bibinfo{pages}{1--15}.
\newblock


\bibitem[Popic et~al\mbox{.}(2020)]%
        {popic2020database}
\bibfield{author}{\bibinfo{person}{Deian Popic}, \bibinfo{person}{Simona~G Pacozzi}, {and} \bibinfo{person}{Corinna~S Martarelli}.} \bibinfo{year}{2020}\natexlab{}.
\newblock \showarticletitle{Database of virtual objects to be used in psychological research}.
\newblock \bibinfo{journal}{\emph{Plos one}} \bibinfo{volume}{15}, \bibinfo{number}{9} (\bibinfo{year}{2020}), \bibinfo{pages}{e0238041}.
\newblock


\bibitem[Prouzeau et~al\mbox{.}(2020)]%
        {prouzeau2020corsican}
\bibfield{author}{\bibinfo{person}{Arnaud Prouzeau}, \bibinfo{person}{Yuchen Wang}, \bibinfo{person}{Barrett Ens}, \bibinfo{person}{Wesley Willett}, {and} \bibinfo{person}{Tim Dwyer}.} \bibinfo{year}{2020}\natexlab{}.
\newblock \showarticletitle{Corsican twin: Authoring in situ augmented reality visualisations in virtual reality}. In \bibinfo{booktitle}{\emph{Proceedings of the international conference on advanced visual interfaces}}. \bibinfo{pages}{1--9}.
\newblock


\bibitem[Roediger~III et~al\mbox{.}(2007)]%
        {roediger2007science}
\bibfield{author}{\bibinfo{person}{Henry~L Roediger~III}, \bibinfo{person}{Yadin Dudai}, {and} \bibinfo{person}{Susan~M Fitzpatrick}.} \bibinfo{year}{2007}\natexlab{}.
\newblock \bibinfo{booktitle}{\emph{Science of memory: Concepts}}.
\newblock \bibinfo{publisher}{Oxford University Press}.
\newblock


\bibitem[Saffo et~al\mbox{.}(2023)]%
        {saffo2023unraveling}
\bibfield{author}{\bibinfo{person}{David Saffo}, \bibinfo{person}{Sara Di~Bartolomeo}, \bibinfo{person}{Tarik Crnovrsanin}, \bibinfo{person}{Laura South}, \bibinfo{person}{Justin Raynor}, \bibinfo{person}{Caglar Yildirim}, {and} \bibinfo{person}{Cody Dunne}.} \bibinfo{year}{2023}\natexlab{}.
\newblock \showarticletitle{Unraveling the design space of immersive analytics: A systematic review}.
\newblock \bibinfo{journal}{\emph{Small}} \bibinfo{volume}{6}, \bibinfo{number}{7} (\bibinfo{year}{2023}), \bibinfo{pages}{4}.
\newblock


\bibitem[Satriadi et~al\mbox{.}(2023)]%
        {satriadi2023proxsituated}
\bibfield{author}{\bibinfo{person}{Kadek~Ananta Satriadi}, \bibinfo{person}{Andrew Cunningham}, \bibinfo{person}{Ross~T Smith}, \bibinfo{person}{Tim Dwyer}, \bibinfo{person}{Adam Drogemuller}, {and} \bibinfo{person}{Bruce~H Thomas}.} \bibinfo{year}{2023}\natexlab{}.
\newblock \showarticletitle{ProxSituated Visualization: An Extended Model of Situated Visualization using Proxies for Physical Referents}. In \bibinfo{booktitle}{\emph{Proceedings of the 2023 CHI Conference on Human Factors in Computing Systems}}. \bibinfo{pages}{1--20}.
\newblock


\bibitem[Satriadi et~al\mbox{.}(2022)]%
        {satriadi2022augmented}
\bibfield{author}{\bibinfo{person}{Kadek~Ananta Satriadi}, \bibinfo{person}{Andrew Cunningham}, \bibinfo{person}{Bruce~H Thomas}, \bibinfo{person}{Adam Drogemuller}, \bibinfo{person}{Antoine Odi}, \bibinfo{person}{Niki Patel}, \bibinfo{person}{Cathlyn Aston}, {and} \bibinfo{person}{Ross~T Smith}.} \bibinfo{year}{2022}\natexlab{}.
\newblock \showarticletitle{Augmented Scale Models: Presenting Multivariate Data Around Physical Scale Models in Augmented Reality}. In \bibinfo{booktitle}{\emph{2022 IEEE International Symposium on Mixed and Augmented Reality (ISMAR)}}. IEEE, \bibinfo{pages}{54--63}.
\newblock


\bibitem[Schmalstieg and Hollerer(2016)]%
        {schmalstieg2016augmented}
\bibfield{author}{\bibinfo{person}{Dieter Schmalstieg} {and} \bibinfo{person}{Tobias Hollerer}.} \bibinfo{year}{2016}\natexlab{}.
\newblock \bibinfo{booktitle}{\emph{Augmented reality: principles and practice}}.
\newblock \bibinfo{publisher}{Addison-Wesley Professional}.
\newblock


\bibitem[Schwabe et~al\mbox{.}(2009)]%
        {schwabe2009stress}
\bibfield{author}{\bibinfo{person}{Lars Schwabe}, \bibinfo{person}{Andreas B{\"o}hringer}, {and} \bibinfo{person}{Oliver~T Wolf}.} \bibinfo{year}{2009}\natexlab{}.
\newblock \showarticletitle{Stress disrupts context-dependent memory}.
\newblock \bibinfo{journal}{\emph{Learning \& Memory}} \bibinfo{volume}{16}, \bibinfo{number}{2} (\bibinfo{year}{2009}), \bibinfo{pages}{110--113}.
\newblock


\bibitem[Shin et~al\mbox{.}(2023)]%
        {shin2023reality}
\bibfield{author}{\bibinfo{person}{Sungbok Shin}, \bibinfo{person}{Andrea Batch}, \bibinfo{person}{Peter~WS Butcher}, \bibinfo{person}{Panagiotis~D Ritsos}, {and} \bibinfo{person}{Niklas Elmqvist}.} \bibinfo{year}{2023}\natexlab{}.
\newblock \showarticletitle{The Reality of the Situation: A Survey of Situated Analytics}.
\newblock \bibinfo{journal}{\emph{IEEE Transactions on Visualization and Computer Graphics}} (\bibinfo{year}{2023}).
\newblock


\bibitem[Shin et~al\mbox{.}(2021)]%
        {shin2021context}
\bibfield{author}{\bibinfo{person}{Yeon~Soon Shin}, \bibinfo{person}{Rolando Mas{\'\i}s-Obando}, \bibinfo{person}{Neggin Keshavarzian}, \bibinfo{person}{Riya D{\'a}ve}, {and} \bibinfo{person}{Kenneth~A Norman}.} \bibinfo{year}{2021}\natexlab{}.
\newblock \showarticletitle{Context-dependent memory effects in two immersive virtual reality environments: On Mars and underwater}.
\newblock \bibinfo{journal}{\emph{Psychonomic Bulletin \& Review}} \bibinfo{volume}{28}, \bibinfo{number}{2} (\bibinfo{year}{2021}), \bibinfo{pages}{574--582}.
\newblock


\bibitem[Smith(1994)]%
        {smith1994theoretical}
\bibfield{author}{\bibinfo{person}{SM Smith}.} \bibinfo{year}{1994}\natexlab{}.
\newblock \showarticletitle{Theoretical principles of context-dependent}.
\newblock \bibinfo{journal}{\emph{Theoretical aspects of memory}} (\bibinfo{year}{1994}), \bibinfo{pages}{168--195}.
\newblock


\bibitem[Smith(2007)]%
        {smith2007context}
\bibfield{author}{\bibinfo{person}{Steven~M Smith}.} \bibinfo{year}{2007}\natexlab{}.
\newblock \showarticletitle{Context: A reference for focal experience}.
\newblock \bibinfo{journal}{\emph{Science of memory: Concepts}} (\bibinfo{year}{2007}), \bibinfo{pages}{111--114}.
\newblock


\bibitem[Smith and Vela(2001)]%
        {smith2001environmental}
\bibfield{author}{\bibinfo{person}{Steven~M Smith} {and} \bibinfo{person}{Edward Vela}.} \bibinfo{year}{2001}\natexlab{}.
\newblock \showarticletitle{Environmental context-dependent memory: A review and meta-analysis}.
\newblock \bibinfo{journal}{\emph{Psychonomic bulletin \& review}} \bibinfo{volume}{8}, \bibinfo{number}{2} (\bibinfo{year}{2001}), \bibinfo{pages}{203--220}.
\newblock


\bibitem[Tan et~al\mbox{.}(2001)]%
        {tan2001infocockpit}
\bibfield{author}{\bibinfo{person}{Desney~S Tan}, \bibinfo{person}{Jeanine~K Stefanucci}, \bibinfo{person}{Dennis~R Proffitt}, {and} \bibinfo{person}{Randy Pausch}.} \bibinfo{year}{2001}\natexlab{}.
\newblock \showarticletitle{The Infocockpit: Providing location and place to aid human memory}. In \bibinfo{booktitle}{\emph{Proceedings of the 2001 workshop on Perceptive user interfaces}}. \bibinfo{pages}{1--4}.
\newblock


\bibitem[Tatzgern et~al\mbox{.}(2014)]%
        {tatzgern2014hedgehog}
\bibfield{author}{\bibinfo{person}{Markus Tatzgern}, \bibinfo{person}{Denis Kalkofen}, \bibinfo{person}{Raphael Grasset}, {and} \bibinfo{person}{Dieter Schmalstieg}.} \bibinfo{year}{2014}\natexlab{}.
\newblock \showarticletitle{Hedgehog labeling: View management techniques for external labels in 3D space}. In \bibinfo{booktitle}{\emph{2014 IEEE Virtual Reality (VR)}}. IEEE, \bibinfo{pages}{27--32}.
\newblock


\bibitem[Thomas et~al\mbox{.}(2018)]%
        {thomas2018situated}
\bibfield{author}{\bibinfo{person}{Bruce~H Thomas}, \bibinfo{person}{Gregory~F Welch}, \bibinfo{person}{Pierre Dragicevic}, \bibinfo{person}{Niklas Elmqvist}, \bibinfo{person}{Pourang Irani}, \bibinfo{person}{Yvonne Jansen}, \bibinfo{person}{Dieter Schmalstieg}, \bibinfo{person}{Aur{\'e}lien Tabard}, \bibinfo{person}{Neven~AM ElSayed}, \bibinfo{person}{Ross~T Smith}, {et~al\mbox{.}}} \bibinfo{year}{2018}\natexlab{}.
\newblock \showarticletitle{Situated Analytics.}
\newblock \bibinfo{journal}{\emph{Immersive analytics}}  \bibinfo{volume}{11190} (\bibinfo{year}{2018}), \bibinfo{pages}{185--220}.
\newblock


\bibitem[Urakami et~al\mbox{.}(2023)]%
        {urakami2023augmenting}
\bibfield{author}{\bibinfo{person}{Jacqueline Urakami}, \bibinfo{person}{Akito Moriwaki}, \bibinfo{person}{Shotaro Nagao}, \bibinfo{person}{Kousuke Osumi}, \bibinfo{person}{Erika Yamamoto}, {and} \bibinfo{person}{Toshikazu Kanaoka}.} \bibinfo{year}{2023}\natexlab{}.
\newblock \showarticletitle{Augmenting Auditory Attention and Memory to Reduce Cognitive Load in Dual Tasks: A wearable device to augment auditory attention and memory to improve performance in a duals task}. In \bibinfo{booktitle}{\emph{Extended Abstracts of the 2023 CHI Conference on Human Factors in Computing Systems}}. \bibinfo{pages}{1--6}.
\newblock


\bibitem[Wallet et~al\mbox{.}(2011)]%
        {wallet2011virtual}
\bibfield{author}{\bibinfo{person}{Gr{\'e}gory Wallet}, \bibinfo{person}{H{\'e}l{\`e}ne Sauz{\'e}on}, \bibinfo{person}{Prashant~Arvind Pala}, \bibinfo{person}{Florian Larrue}, \bibinfo{person}{Xia Zheng}, {and} \bibinfo{person}{Bernard N'Kaoua}.} \bibinfo{year}{2011}\natexlab{}.
\newblock \showarticletitle{Virtual/real transfer of spatial knowledge: Benefit from visual fidelity provided in a virtual environment and impact of active navigation}.
\newblock \bibinfo{journal}{\emph{Cyberpsychology, Behavior, and Social Networking}} \bibinfo{volume}{14}, \bibinfo{number}{7-8} (\bibinfo{year}{2011}), \bibinfo{pages}{417--423}.
\newblock


\bibitem[W{\"a}lti et~al\mbox{.}(2019)]%
        {walti2019reinstating}
\bibfield{author}{\bibinfo{person}{Michel~Juhani W{\"a}lti}, \bibinfo{person}{Daniel~Graham Woolley}, {and} \bibinfo{person}{Nicole Wenderoth}.} \bibinfo{year}{2019}\natexlab{}.
\newblock \showarticletitle{Reinstating verbal memories with virtual contexts: Myth or reality?}
\newblock \bibinfo{journal}{\emph{PloS one}} \bibinfo{volume}{14}, \bibinfo{number}{3} (\bibinfo{year}{2019}), \bibinfo{pages}{e0214540}.
\newblock


\bibitem[Wen et~al\mbox{.}(2022)]%
        {wen2022effects}
\bibfield{author}{\bibinfo{person}{Zhen Wen}, \bibinfo{person}{Wei Zeng}, \bibinfo{person}{Luoxuan Weng}, \bibinfo{person}{Yihan Liu}, \bibinfo{person}{Mingliang Xu}, {and} \bibinfo{person}{Wei Chen}.} \bibinfo{year}{2022}\natexlab{}.
\newblock \showarticletitle{Effects of view layout on situated analytics for multiple-view representations in immersive visualization}.
\newblock \bibinfo{journal}{\emph{IEEE Transactions on Visualization and Computer Graphics}} \bibinfo{volume}{29}, \bibinfo{number}{1} (\bibinfo{year}{2022}), \bibinfo{pages}{440--450}.
\newblock


\bibitem[White and Feiner(2009)]%
        {white2009sitelens}
\bibfield{author}{\bibinfo{person}{Sean White} {and} \bibinfo{person}{Steven Feiner}.} \bibinfo{year}{2009}\natexlab{}.
\newblock \showarticletitle{SiteLens: situated visualization techniques for urban site visits}. In \bibinfo{booktitle}{\emph{Proceedings of the SIGCHI conference on human factors in computing systems}}. \bibinfo{pages}{1117--1120}.
\newblock


\bibitem[White et~al\mbox{.}(2006)]%
        {white2006virtual}
\bibfield{author}{\bibinfo{person}{Sean White}, \bibinfo{person}{Steven Feiner}, {and} \bibinfo{person}{Jason Kopylec}.} \bibinfo{year}{2006}\natexlab{}.
\newblock \showarticletitle{Virtual vouchers: Prototyping a mobile augmented reality user interface for botanical species identification}. In \bibinfo{booktitle}{\emph{3D User Interfaces (3DUI'06)}}. IEEE, \bibinfo{pages}{119--126}.
\newblock


\bibitem[White(2009)]%
        {white2009interaction}
\bibfield{author}{\bibinfo{person}{Sean~Michael White}.} \bibinfo{year}{2009}\natexlab{}.
\newblock \bibinfo{booktitle}{\emph{Interaction and presentation techniques for situated visualization}}.
\newblock \bibinfo{publisher}{Columbia University}.
\newblock


\bibitem[Whitlock et~al\mbox{.}(2020)]%
        {whitlock2020hydrogenar}
\bibfield{author}{\bibinfo{person}{Matt Whitlock}, \bibinfo{person}{Danielle~Albers Szafir}, {and} \bibinfo{person}{Kenny Gruchalla}.} \bibinfo{year}{2020}\natexlab{}.
\newblock \showarticletitle{HydrogenAR: Interactive Data-Driven Presentation of Dispenser Reliability}. In \bibinfo{booktitle}{\emph{2020 IEEE International Symposium on Mixed and Augmented Reality (ISMAR)}}. IEEE, \bibinfo{pages}{704--712}.
\newblock


\bibitem[Willett et~al\mbox{.}(2016)]%
        {willett2016embedded}
\bibfield{author}{\bibinfo{person}{Wesley Willett}, \bibinfo{person}{Yvonne Jansen}, {and} \bibinfo{person}{Pierre Dragicevic}.} \bibinfo{year}{2016}\natexlab{}.
\newblock \showarticletitle{Embedded data representations}.
\newblock \bibinfo{journal}{\emph{IEEE transactions on visualization and computer graphics}} \bibinfo{volume}{23}, \bibinfo{number}{1} (\bibinfo{year}{2016}), \bibinfo{pages}{461--470}.
\newblock


\bibitem[Wobbrock et~al\mbox{.}(2011)]%
        {wobbrock2011aligned}
\bibfield{author}{\bibinfo{person}{Jacob~O Wobbrock}, \bibinfo{person}{Leah Findlater}, \bibinfo{person}{Darren Gergle}, {and} \bibinfo{person}{James~J Higgins}.} \bibinfo{year}{2011}\natexlab{}.
\newblock \showarticletitle{The aligned rank transform for nonparametric factorial analyses using only anova procedures}. In \bibinfo{booktitle}{\emph{Proceedings of the SIGCHI conference on human factors in computing systems}}. \bibinfo{pages}{143--146}.
\newblock


\bibitem[Zheng et~al\mbox{.}(2022)]%
        {zheng2022stare}
\bibfield{author}{\bibinfo{person}{Mengya Zheng}, \bibinfo{person}{Xingyu Pan}, \bibinfo{person}{Nestor~Velasco Bermeo}, \bibinfo{person}{Rosemary~J Thomas}, \bibinfo{person}{David Coyle}, \bibinfo{person}{Gregory~MP O’hare}, {and} \bibinfo{person}{Abraham~G Campbell}.} \bibinfo{year}{2022}\natexlab{}.
\newblock \showarticletitle{Stare: Augmented reality data visualization for explainable decision support in smart environments}.
\newblock \bibinfo{journal}{\emph{IEEE Access}}  \bibinfo{volume}{10} (\bibinfo{year}{2022}), \bibinfo{pages}{29543--29557}.
\newblock


\end{thebibliography}

\appendix
\newpage
\section{Appendix}

\subsection{Context-dependent memory concepts}
\label{appendixcdm}
Relevant introductory concepts of CDM studies adapted from existing literature~\cite{smith1994theoretical, roediger2007science, smith2001environmental}.

\subsubsection{Paradigms} 
Paradigms, also called effects, are the patterns for observation of CDM. Three known paradigms are context reinstatement, interference reduction, and multiple contexts.

\begin{figure}[htb!]
    \centering
    \includegraphics[width=0.7\linewidth]{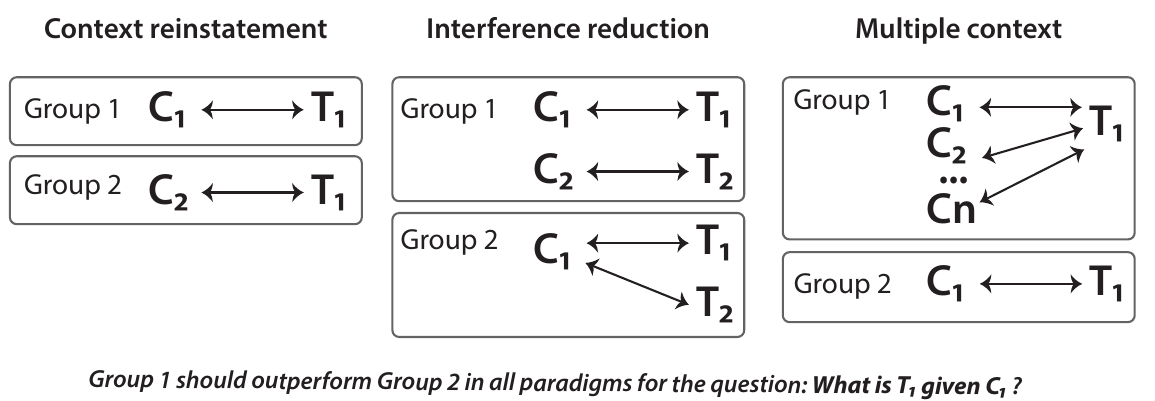}
    \caption{Context reinstatement, interference reduction, and multiple contexts are common CDM effects paradigms or effects. C is context and T is target information. }
   \label{fig:paradigms}
\end{figure}

\textbf{Context Reinstatement Effect}: This phenomenon enhances memory performance for cases where the learning context is re-instantiated during the recall. Suppose there are two groups of people. Group 1 learns information T$_{1}$ paired with context C$_{1}$ while Group 2 learns the same information but paired with C$_{2}$. When they are asked to recall T$_{1}$ given the presence of C$_{1}$, Group 1's performance is better because the presence of C$_{1}$ helps them memorize the information (Figure \ref{fig:paradigms}-left).

\textbf{Interference Reduction effect}: This effect leads to enhanced memory performance when there is reduced interference between context and target information. For instance, consider two groups: Group 1 learns two pairs of context-target, C${1}$ and T${1}$, and C${2}$ and T${2}$. Group 2, on the other hand, learns both T${1}$ and T${2}$ with the same context, C${1}$. When both groups are tasked with the same memory recall, Group 1 outperforms Group 2. This is because in Group 1, there is a clear association between context and target, while in Group 2, the presence of multiple target information (T${1}$, T$_{2}$) leads to interference between them.

\textbf{Multiple Contexts Effect}: This effect enhances memory performance when the target information is learned in multiple contexts. To illustrate, let's consider two groups: Group 1 learns the target T${1}$ in multiple contexts, denoted as C${1}$, C${2}$, C${3}$, ..., C${n}$, while Group 2 exclusively learns T${1}$ in the C$_{1}$ context. The Multiple Contexts effect becomes evident when Group 1 outperforms Group 2. This phenomenon is attributed to the idea that Group 1's learning process encodes more contextual cues compared to Group 2, resulting in enhanced memory recall.

\subsubsection{Overshadow and outshining}

Overshadowing and outshining are two significant concepts in Context-Dependent Memory (CDM) that relate to the strength of cues associated with multiple features during both the learning and recall stages.

In the learning phase, overshadowing occurs when two context features, such as C${1}$ and C${2}$, are linked to a target memory. Consider three scenarios to illustrate this concept.
\begin{itemize}
    \item     In Scenario 1, the target memory is learned in the presence of C${1}$, and during the recall phase, C${1}$ is provided as the context. In this case, memory performance is labeled as X.
     \item
     In Scenario 2, the same target is learned with both C${1}$ and C${2}$ present simultaneously. However, during the recall phase, only C${2}$ (instead of C${1}$) is presented as the context. Memory performance in this scenario is X+1, indicating an improvement compared to C$_{1}$ alone.
\item
    In Scenario 3, under the same learning conditions as Scenario 2, the context provided during recall is C${1}$ (rather than C${2}$). In this case, memory performance is X-1.
\end{itemize}

\begin{figure*}[htb!]
    \centering
    \includegraphics[width=0.9\linewidth]{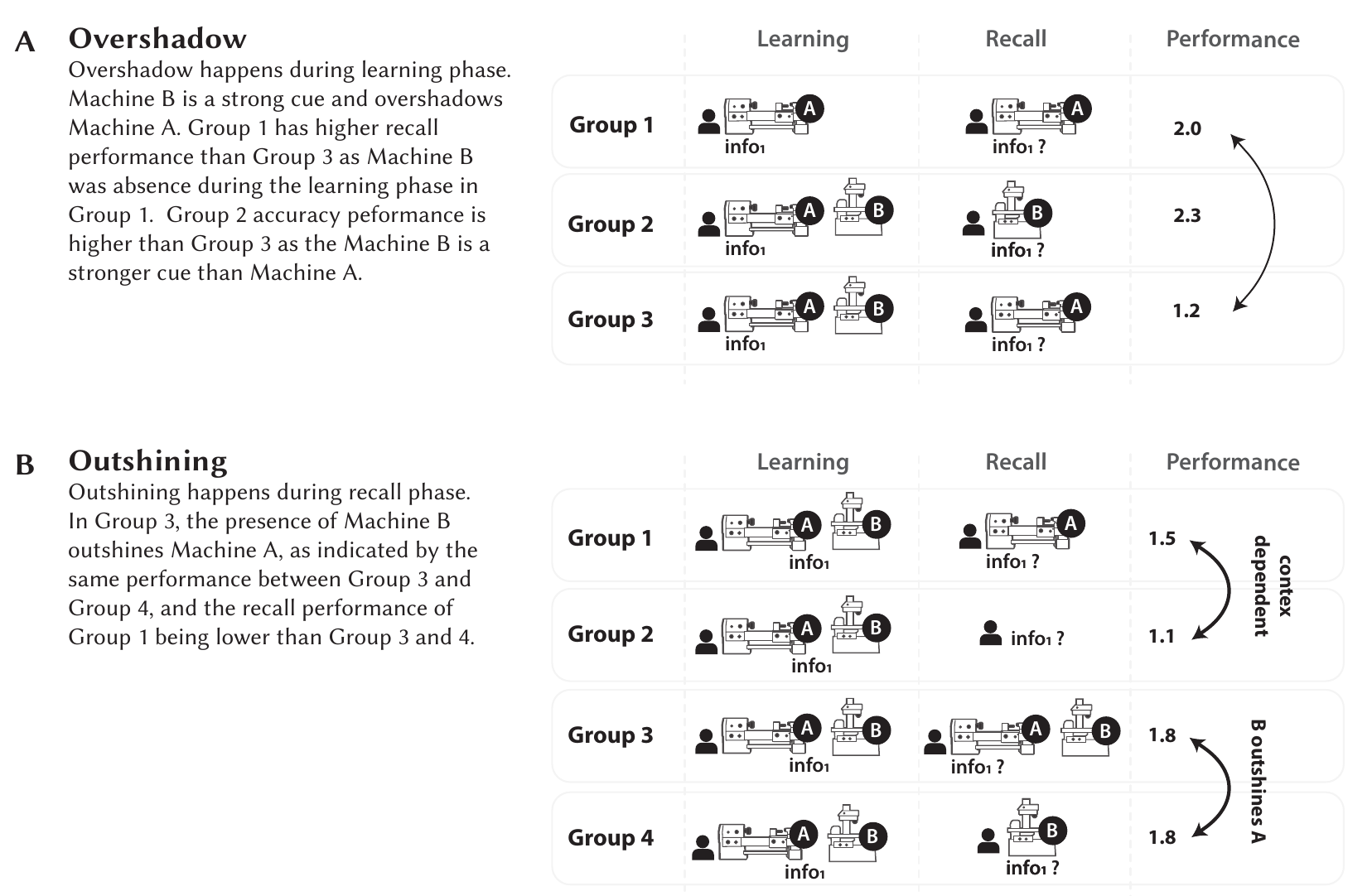}
    \caption{Illustrations and descriptions of overshadow and outshining, adapted from~\cite{smith1994theoretical}.}
   \label{fig:cdm_concepts}
\end{figure*}

Here, C${2}$ is considered a stronger cue than C${1}$ because it overshadows C$_{1}$ during the learning stages of Scenarios 2 and 3. While overshadowing occurs during the learning phase, outshining is a similar phenomenon that happens during the recall phase. It occurs when multiple features are presented during the learning stage, and the contextual cue of one feature is hindered by another during recall. For a visual representation, please refer to Figure~\ref{fig:cdm_concepts}.

\subsection{Related work}
\label{appendixrelatedwork}
This section contains related work shown in Table \ref{table:lit}.
\begin{table*}[]
\footnotesize
\caption{Recent relevant studies  ranging from the year 2018 to 2023, sorted from the oldest to the most recent. LI: Learned Information, RT: Retention Time, RM: Recall Method, LT: Learning Time.}
\label{table:lit}
\begin{tblr}{p{2cm}p{1cm}p{3cm}p{1cm}p{1cm}p{2cm}p{4cm}}
\textbf{Study}  & \textbf{Display} & \textbf{Context} & \textbf{LI}   & \textbf{{RT}} &  \textbf{RM} & \textbf{CDM Evidence}                                            \\
\hline

de Back et al.~\cite{de2018applicability} & 
CAVE & 
{Panoramic Mountain, Panoramic Underwater} & 
words & 
immediate & 
recognition  & 
A significant effect was found between Mountain-Mountain and Mountain-Underwater conditions but no difference was found between Underwater-Underwater and Underwater-Mountain conditions.\\

Walti et al.~\cite{walti2019reinstating}\\(Study 3.2) & 
VR & 
{Virtual Environments} & 
words  & 
immediate & 
free recall  & 
No significant effect. \\

Walti et al.~\cite{walti2019reinstating} \\(Study 3.2) &
Desktop & 
{Virtual Environments Pictures}  & 
words & 
immediate      & 
free recall  & 
No significant effect.                                             \\

Walti et al.~\cite{walti2019reinstating} \\(Study 2.1)     & 
 Desktop          & {Environment Pictures}   & 
 words & 
 immediate & 
 free recall   & 
 No significant effect.   \\
 
Walti et al.~\cite{walti2019reinstating} \\(Study 2.2) & 
VR & 
{Virtual Environments} & 
words  & 
immediate  & 
free recall & 
No significant effect.  \\

Shin et al.~\cite{shin2021context} \\(long-term recall) & 
VR & 
{VE Mars, VE Underwater } & reading words  & 
24 hours   & {cued recall\\(2 min)}  & 
Significant effect.                                               \\

Shin et al.~\cite{shin2021context} \\(short-term recall) & 
VR & 
{VE Mars, VE Underwater} & 
words & 
immediate  & {cued recall \\(2 min)} & 
No significant effect. 
\\

Parker et al.~\cite{parker2020exploring}  & Real, VR         & 
VE Underwater, Real world& sentences, words, pictures
& immediate
& listening and recognition 
 & No significant effect.\\

Lamers and Lanen~\cite{lamers2021changing}  & Real, VR         & Real study space, Virtual replica of the space 
 & sentences, words, pictures
& 24 hours

& free recall, cued recall, recognition 
& Significant effect (17\% differences). Recall accuracy is 24\% lower when items were memorized in VR when compared to memorization in a real environment. \\

Essoe et al.~\cite{essoe2022enhancing}  & 
VR & 
{VE Fantasy (single vs dual-context)} & 
words from 2 languages & 
24 hours, one week  & free recall & 
Significant interference reduction effect. Unique context-information mapping shows reduced interference and improved one-week retention by approx. 16\%. These results are only apparent if the VR environments are experienced as real environments. 
\\

Chochol{\'a}{\v{c}}kov{\'a} et al.~\cite{chocholavckova2023context}  & 
VR & 
Indoor VE, Outdoor VE  & 
words & 
immediate  &  free recall & 
No significant effect. 
\\

Mizuho et al.~\cite{mizuho2023virtual}  & 
Real, VR & 
Real, Realistic virtual replica, Simple virtual replica & 
words & 
immediate  &  free recall & 
No significant effect. 
\\

\end{tblr}
\end{table*}

\subsection{Design space description}
\label{appendixdesignspace}

This section contains the description of the design space and illustrations shown in Figure \ref{fig:design_space} and Figure \ref{fig:examples}.

\begin{enumerate}[leftmargin=*, label=(\Alph*)]
    \item \textit{Immediate Situated, environment referent.}
In this category, the referent is a space in the physical world, meaning the context of the data is the immediate environment around the user (A in Figure \ref{fig:examples} and \ref{fig:design_space}). Existing work related to this category includes ~\cite{white2009sitelens, moere2012designing}.

\item \textit{Immediate Situated, a single referent.} 
In this category, the referent is an object in the environment rather than space. For example, a platypus in an exhibit is surrounded by the environment (B in Figure \ref{fig:examples} and \ref{fig:design_space}). Existing work related to this category includes~\cite{collins2020augmented}.

\item \textit{Immediate Situated, homogeneous referents.}
In this category, multiple objects that have similar features in the physical world are the referents. For example, imagine a savana tour where a pack of zebras are considered referent (C in Figure \ref{fig:examples} and \ref{fig:design_space}). Existing work related to this category includes~\cite{guarese2020augmented}.

\item \textit{Immediate Situated, heterogenous referents}. 
In this category, multiple objects that have distinctive features in the physical world are the referents.
 For example, consider a situated visualizaiton in a classroom where the students are the referents. The students are likely to have unique features to remember (E in Figure \ref{fig:examples} and \ref{fig:design_space}).
Existing work related to this category includes use cases in ~\cite{fleck2022ragrug}.

\item \textit{Non-VR ProxSituated, environment referent.}
In this category, the referent is a space in the physical world but it is presented as proxies in the user's workspace. For example, imagine a data visualization on a replica of a city (E in Figure \ref{fig:examples} and \ref{fig:design_space}). Existing work related to this category includes~\cite{alonso2018cityscope}.

\item \textit{Non-VR ProxSituated, a single referent.}
In this category, the referent is an object in the physical world but it is presented as proxies in the user's workspace. For example, imagine a data visualization of a mining machine presented on a desktop screen (F in Figure \ref{fig:examples} and \ref{fig:design_space}). Existing work related to this category includes ~\cite{tatzgern2014hedgehog} where a scale model of a person's head is used as a referent, or \cite{pooryousef2023working} where a virtual body is used as proxy referent in augmented reality. 

\item \textit{Non-VR ProxSituated, homogeneous referents.}
In this category, the referents are multiple homogenous objects in the physical world but they are presented as proxies in the user's workspace. For example, imagine a data visualization of wind turbines presented on a desktop screen (G in Figure \ref{fig:examples} and \ref{fig:design_space}). 

\item \textit{Non-VR ProxSituated, heterogeneous referents.}
In this category, the referents are multiple heterogeneous objects in the physical world but they are presented as proxies in the user's workspace. For example, data visualization on different buildings presented on a desktop screen as illustrated in Figure \ref{fig:examples}-H. Existing work related to this category includes~\cite{lin2022quest, wen2022effects, satriadi2022augmented, ens2020uplift}.

\item \textit{VR ProxSituated, environment referent.}
In this category, the referent is a space in the physical world but it is presented in real-scale proxies in VR. For example, immersive geovisualization of bush fire data (I in Figure \ref{fig:examples} and \ref{fig:design_space}). 

\item \textit{VR ProxSituated, a single referent.}
In this category, the referent is an object in the physical world but it is presented in real-scale proxies in VR. For example, immersive visualization of a statue (J in Figure \ref{fig:examples} and \ref{fig:design_space}). Existing work related to this category includes~\cite{prouzeau2020corsican}.

\item \textit{VR ProxSituated, homogeneous referents.}
In this category, the referents are homogeneous objects in the physical world but they are presented in real-scale proxies in VR. For example, multiple laboratory equipment of the same type presented in VR (K in Figure \ref{fig:examples} and \ref{fig:design_space}).

\item \textit{VR ProxSituated, heterogeneous referents.}
In this category, the referents are heterogeneous objects in the physical world but they are presented in real-scale proxies in VR. For example, data visualization around different industrial vehicles in a construction site (L in Figure \ref{fig:examples} and \ref{fig:design_space}). Existing work related to this category includes~\cite{zheng2022stare} (the VR simulation).

\end{enumerate}

\end{document}